\documentclass[onecolumn,preprintnumbers,amsmath,amssymb]{revtex4}

\usepackage{graphicx}
\usepackage[latin1]{inputenc}
\usepackage{dcolumn}
\usepackage{bm}
\usepackage{epsfig}
\usepackage[english]{babel}
\usepackage{xcolor}
\usepackage{calrsfs}

\newtheorem{definition}{Definition}[section]

\begin{document}


\title{The arrow of time across five centuries of classical music}

\author{Alfredo Gonz\'alez-Espinoza$^1$, Gustavo Mart\'inez-Mekler$^{2,3}$, Lucas Lacasa$^{4}$}
\email{l.lacasa@qmul.ac.uk}
\affiliation{$^1$Department of Biology, University of Pennsylvania, Philadelphia, PA 19104 (US)\\
$^2$Instituto de Ciencias F\'isicas, Universidad Nacional Aut\'onoma de Mexico, Cuernavaca (Mexico)\\
$^3$Centro de Ciencias de la Complejidad, Universidad Nacional Aut\'onoma de Mexico, Ciudad de M\'exico (Mexico)\\
$^4$School of Mathematical Sciences, Queen Mary University of London, Mile End Road E14NS, London (UK)\\}%


\begin{abstract}
The concept of time series irreversibility --the degree by which the statistics of signals are not invariant under time reversal-- naturally appears in non-equilibrium physics in stationary systems which operate away from equilibrium and produce entropy. This concept has not been explored to date in the realm of musical scores as these are typically short sequences whose time reversibility estimation could suffer from strong finite size effects which preclude interpretability.
Here we show that the so-called Horizontal Visibility Graph method --which recently was shown to quantify such statistical property even in non-stationary signals-- is a method that can estimate time reversibility of short symbolic sequences, thus unlocking the possibility of exploring such properties in the context of musical compositions.
Accordingly, we analyse over 8000 musical pieces ranging from the Renaissance to the early Modern period and certify that, indeed, most of them display clear signatures of time irreversibility. Since by construction stochastic processes with a linear correlation structure (such as $1/f$ noise) are time reversible, we conclude that musical compositions have a considerably richer structure, that goes beyond the traditional properties retrieved by the power spectrum or similar approaches. We also show that musical compositions display strong signs of nonlinear correlations, that nonlinearity is correlated to irreversibility, and that these are also related to asymmetries in the abundance of musical intervals, which we associate to the narrative underpinning a musical composition. These findings provide tools for the study of musical periods and composers, as well as criteria related to music appreciation and cognition.
\end{abstract}

 \maketitle

\email{l.lacasa@qmul.ac.uk}

\section{Introduction}

The quantitative description of the structure in musical compositions has a long history of interdisciplinary research, with contributions from musical theory, information theory, mathematics to physics.
Traditional quantitative analysis of the temporal structure underlying musical pieces have mainly addressed linear correlations, such as the ones captured by spectral (Fourier) analysis, starting from the pioneering work of Voss and Clarke \cite{Voss1, Voss2} and subsequently followed by a wealth of more in-depth analysis \cite{pnas_ritmo, levitin2015, Hennig2011, Hennig2014, gunduz2005, Telesca2012, Dagdug2007, Jafari2007, Jennings2004, Liu2013, LiuX2010}. It is nowadays widely accepted that music presents a so-called $1/f$ power spectrum and that this is a fingerprint of ``appealing sound" \cite{Schroeder1991, West1990, Gilden1995, Press1978, Bak1996, Kauffman1995}. However, recent evidence has challenged this vision, as it has been suggested that pleasantness could be also related to nonlinearities present in music compositions, a property which by definition is not captured in the power spectra \cite{gustavo1}. These findings motivate further exploration into quantitative ways of measuring structure in music compositions that goes beyond linear theories. Amongst others, musicians and musicologists have addressed the breakdown of continuity and temporality \cite{Hastey}, the effect of asymmetry in melody \cite{VosTroost} and the importance of musical irreversibility \cite{mgrant}. In all cases, a relationship with pleasantness has been explored.\\

\noindent Inspired by both statistical physics and nonlinear dynamical concepts, here we introduce statistical measures for irreversibility, nonlinearity and asymmetry and present a description of music scores by gauging their interrelations. We primarily explore to which extent classical music manifests statistical time irreversibility and further on introduce nonlinearity and asymmetry. Simply put, a stationary signal is (statistically) time reversible if the statistical properties of the signal are invariant under time reversal \cite{Weiss}, whereas  the signal is statistically irreversible in the opposite case. For instance, white noise is a stochastic process known to be statistically reversible: if one listens to temporal white noise and subsequently to the same signal after time reversal, it is not possible to distinguish not only which is which, but also whether they sound different.\\
The notion of statistical time irreversibility, which has been associated with an ``arrow of time", has deep relations in non-equilibrium physics with concepts such as dissipation and entropy production. For instance, the amount of entropy produced by a thermodynamic system out of equilibrium has been linked, in non-equilibrium steady states (NESS), to the extent in this system displays time irreversibility \cite{Roldan1, Roldan2, Roldan11}.
Interestingly, it is well known that a large family of stochastic processes --which include pink or $1/f$ noise as a special case-- are statistically reversible \cite{kennel}. The presence or absence of statistical reversibility is a priori a well-suited concept to explore to which extent the regularities and patterns present in musical compositions go beyond linear correlation structures.\\
Standard methods that estimate irreversibility in (discrete) stationary signals usually require long time series sizes for an accurate estimation when the alphabet (number of different states) is large, simply because the amount of possible $m$-grams scales exponentially with the alphabet size. This is a problem expected to emerge in musical compositions, as these are seldom large, typically consisting of sequences of some hundreds of notes, and the alphabet size (e.g. the number of different notes involved in the piece) is rarely exponentially small. Here we leverage on a recently introduced approach, {\it horizontal visibility graph irreversibility} (HVG-I) \cite{EPJB2012, PRE2015}, which we argue actually bridges this gap and is able to extract meaningful measures of statistical irreversibility in short sequences and applies to both stationary and non-stationary signals. We extend the method to deal with short sequences by defining a measure of HVG-irreversibility --which we can link to entropy production-- and a confidence index which states when a certain irreversibility value is genuine or, on the contrary, is just a finite size effect and therefore spurious. Equipped with these tools, we can then explore time series irreversibility in music. Interestingly, we find that a large amount of compositions indeed display time irreversibility and therefore can be understood as signals generated by systems which operate out of equilibrium and producing entropy. We explore in detail the relation between irreversibility, entropy production and the presence of nonlinear temporal correlations in music, establishing that irreversibility is a key feature of musical compositions which is not related to linear information displayed by the power spectrum, and we are finally able to interpret such fingerprints in terms of musical composition.\\

\noindent The rest of the paper is as follows: in section II we present the database, which consists of over $8000$ musical compositions from $77$ different composers spanning several centuries and different musical periods, from the Renaissance to the beginning of the Modern Period. In section III we present the theory and methods used to estimate time irreversibility and entropy production in music. In section IV we present the results from these methods and complement them with additional characterisation provided for nonlinearity and interval asymmetry. The former attempts to quantify the temporal correlation structure which persists in musical compositions once linear temporal correlations are removed, whereas the latter is a strongly musical-based notion. We provide a global picture by comparing the performance and relations between irreversibility, nonlinearity and interval asymmetry, and in section V we conclude.

\section{Dataset}
\begin{figure}[h]
    \centering
    \includegraphics[width=0.35\textwidth]{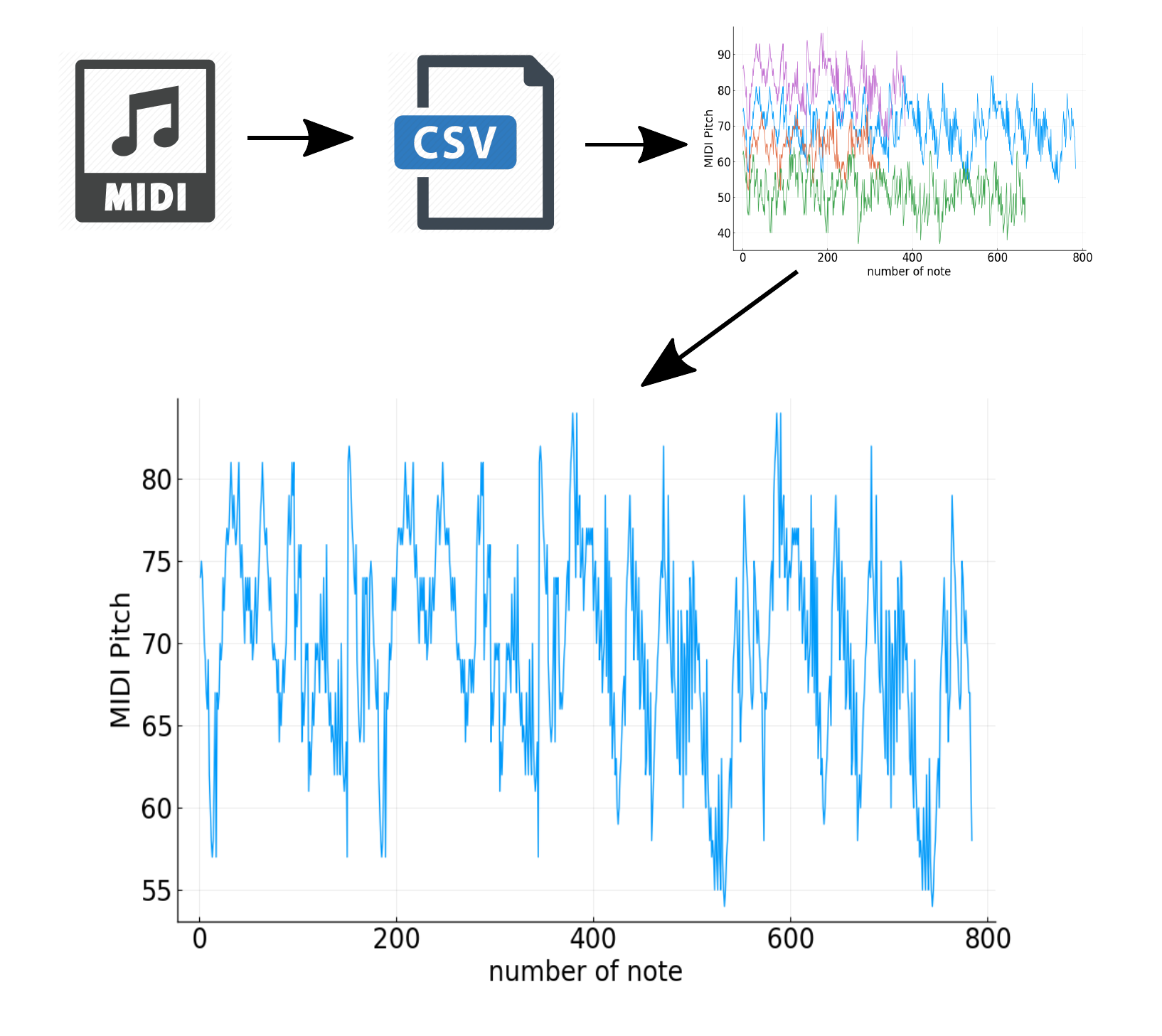}
    \includegraphics[width=0.35\columnwidth]{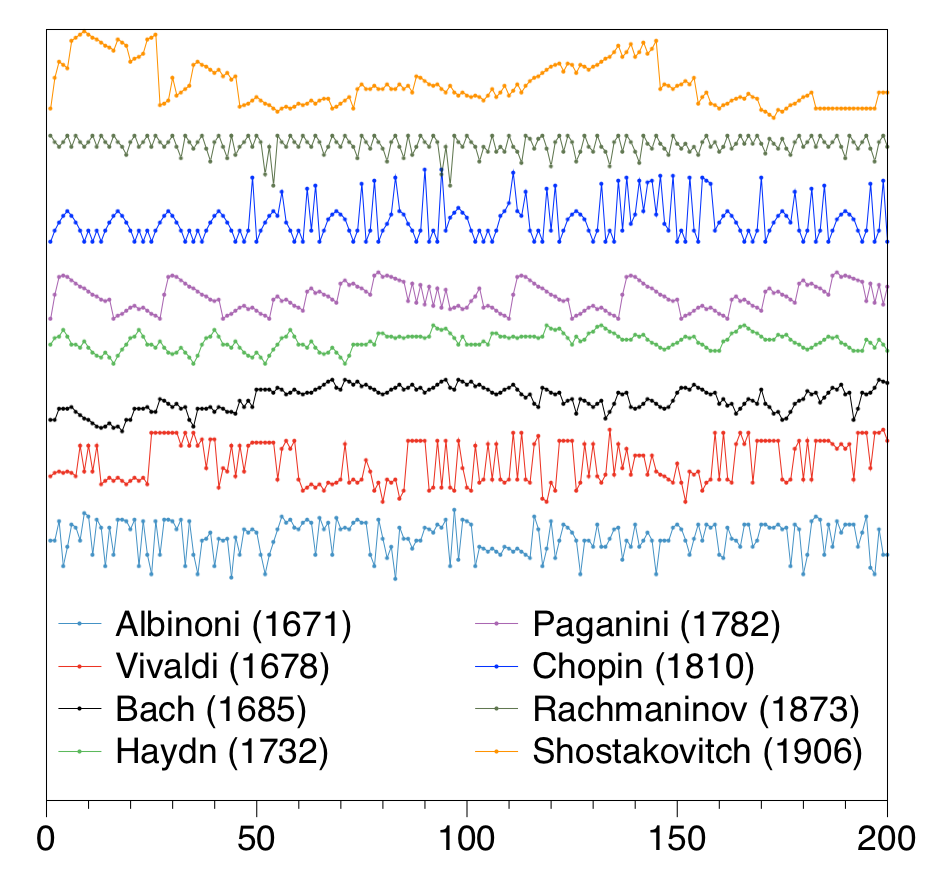}
    \caption{{\bf Data extraction protocol. }(Left panel) Scheme for the process of extracting one pitch sequence from each MIDI file. (Right panel) Samples of pieces from different authors (note that the y-axis has been shifted for illustration purposes).}
    \label{fig:midi}
\end{figure}

We collected 8856 \textsc{midi} files of 77 different composers extracted from the {\em Kunst der fugue} \textsc{midi} dataset \cite{kdf}. Since each piece usually incorporates different voices (different time series, each one corresponding to a different voice), a first task was to decide which of these voices should be considered as the main voice or {\it pitch sequence} of the piece (note that the full, multivariate analysis could be done as well, and we leave that approach for a future work).
To choose the pitch sequence in multi-voice pieces, we used the following two-step criterion: 1) the pitch sequence must be longer than 30 notes and 2) the sequence should have the largest number of different notes from all sequences in the same piece. With this criterion we are assuming that the selected pitch sequence would have most of the relevant features in the musical piece (such as the melody or the theme with its variations). We checked that applying the criterion above was unambigious, and only one sequence could be extracted from each piece.
We processed each \textsc{midi} file by parsing it into a comma separated value (CSV) \cite{midicsv} format and with the aid of a Julia script \cite{age_2019_3242784} we extracted the pitch sequence (see left panel of Fig.\ref{fig:midi} for an illustration).  Some examples of pitch sequences extracted from individual pieces are depicted for illustration in the right panel of Figure \ref{fig:midi}.\\

\begin{figure}[h]
    \centering
    \includegraphics[width=0.35\textwidth]{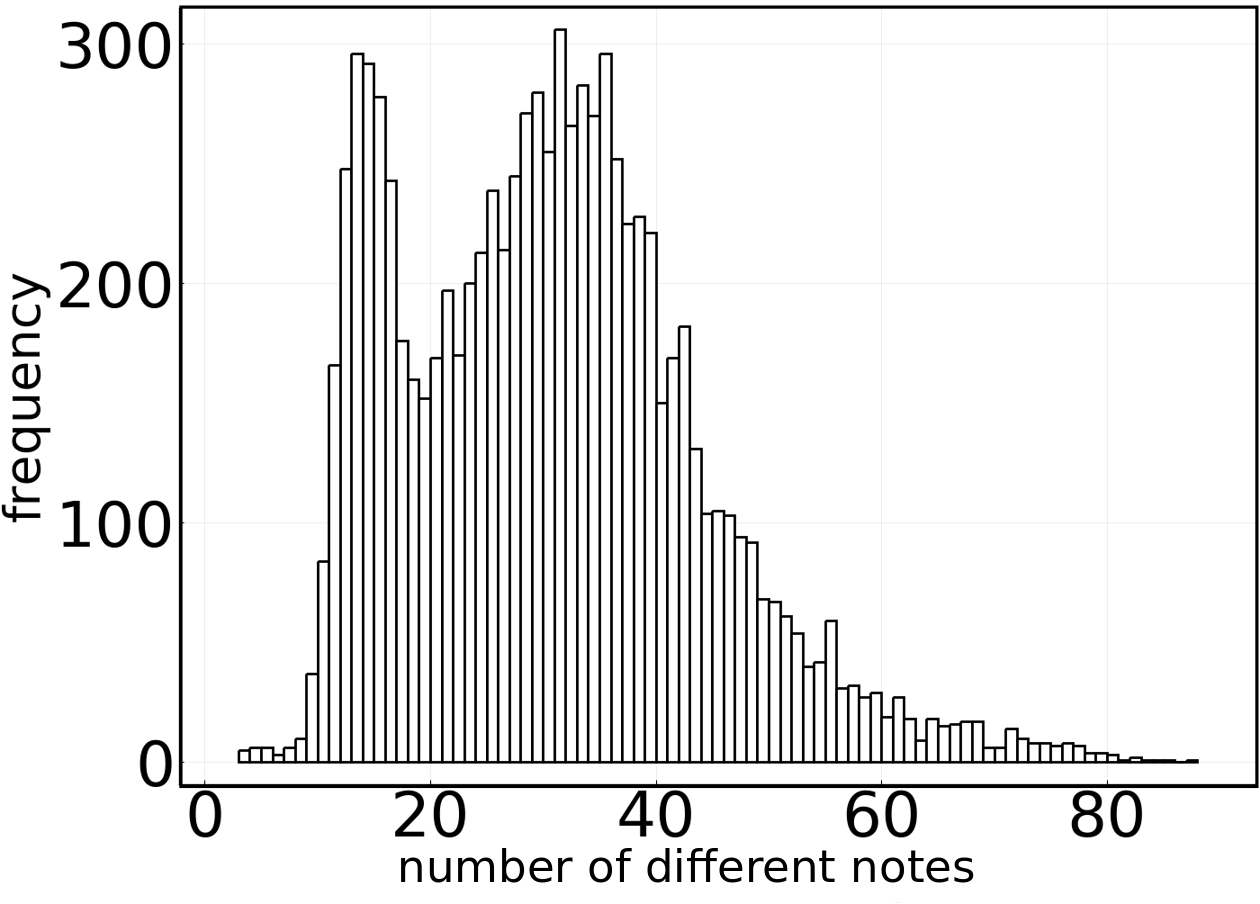}
    \includegraphics[width=0.38\columnwidth]{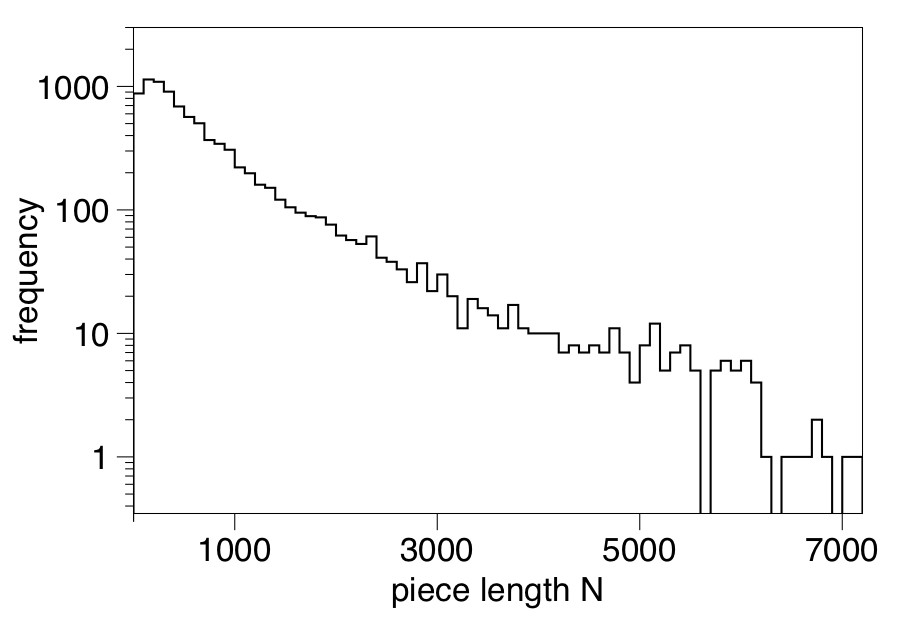}
    \caption{{\bf Alphabet and sequence size. } (Left panel) Histogram of the alphabet (number of different notes) for all the pieces analysed (median $=30$ and kurtosis $=0.6$). (Right panel) Semi-log histogram describing the amount of pieces that have a certain size $N$ (in number of notes). The majority of pieces have less than $10^3$ time steps (median$=456$, kurtosis$=23$). The smallest piece has $N=35$ data points, whereas the largest has $N=15795$ data points.}
    \label{fig:size}
\end{figure}

\noindent To have a rough idea of the characteristics of these sequences, in the right panel of Figure \ref{fig:size} we display the sequence size histogram, whereas in the left panel of the same figure we depict the distribution of the number of different notes (i.e. the alphabet) per piece. As expected, typically the alphabet is too large (in principle there are 128 symbols, and in practice the average alphabet size is 30) and the sequence size too short (median 427) for {\it standard} time irreversibility methods --such as comparing the frequency of $n$-grams in the forward and backward sequence-- to be applicable, hence motivating the use of graph-theoretic approaches such as the one described in the next section.

\section{Time Irreversibility}
\subsection{Reversibility and entropy production in stationary systems}
A stationary process is said to be time reversible if the joint probability distribution of the forward and backward process are statistically equivalent. More concretely, let ${\cal S}=(x_1,x_2,\dots,x_N)$ be a time series of $N$ data, and denote ${\cal S}^*=(x_N,x_{N-1},\dots,x_1)$ the backward time series. The forward and backwards joint distributions are denoted respectively ${\cal P}_F(N):= P(x_1,x_2,\dots,x_N)$, ${\cal P}_B(N):=P(x_N,x_{N-1},\dots,x_1)$. We say that the time series ${\cal S}$ is statistically time reversible if and only if
${\cal P}_F(m) {\buildrel d \over =} {\cal P}_B(m), \forall m=1,\dots,N$, where ${\buildrel d \over =}$ should be interpreted here as having two distributions which cannot be distinguished.
Accordingly, statistical time reversibility is usually known as the property of a time series whose statistics remain the same when the series is flipped. Note that in practice ${\cal P}_F(m\gg1)$ are hard to estimate, and in the event that only a single realisation $\cal S$ is available, ${\cal P}_F(N)$ cannot be estimated at all. In those cases it is customary to estimate ${\cal P}_F(m)$ and ${\cal P}_B(m)$ for  $2\leq m\ll N$, since a sufficient condition for rejecting time reversibility is to reject indistinguishability for small $m$.\\
Gaussian linear processes such as white noise or colored noise, or conservative chaotic processes such as Hamiltonian chaos are statistically time reversible, and related to processes in thermodynamic equilibrium in statistical physics. Nonlinear stochastic processes, or dissipative chaotic processes on the other hand are generally found to be irreversible \cite{kennel}, and are associated to processes that operate away from equilibrium in a thermodynamic sense.\\
There are various possible approaches to quantify the degree of irreversibility, starting from the obvious choice of comparing the $m$-gram statistics in $\cal S$ and ${\cal S}^*$, to more exotic approaches \cite{kennel, Roldan3, EPJB2012}. A notable result dictates that when the signal $x(t)=x_t$ is generated by an underlying thermodynamic system, then the amount of time irreversibility of an (infinitely long, i.e. $N\to \infty$) trajectory $\cal S$ is related to the amount of entropy that the underlying thermodynamic system is producing \cite{Roldan2}. In particular, in the event that all active (i.e. out of equilibrium) degrees of freedom are characterised in the phase-space variable $x$, then the steady-state rate of entropy production $\sigma_{\text{tot}}$ is related to the time irreversibility of $\cal S$ via
\begin{equation}
\sigma_{\text{tot}}=k_B \lim_{m \to \infty} \frac{1}{m} \text{KLD}({\cal P}_F(m) || {\cal P}_B(m)),
\label{entropy}
\end{equation}
where $k_B$ is the Boltzmann constant and KLD$(\cdot ||  \cdot)$ is the Kullback-Leibler divergence. Note that for any distributions $Q$ and $R$,  $\text{KLD}(Q||R)$ defined by: $$\text{KLD}(Q||R)=\sum_x Q(x) \log[Q(x)/R(x)],$$
quantifies their distinguishability since $\text{KLD}(Q||R)=0$ if and only if $Q$ and $R$ are identical, and is positive otherwise. Note also that when the observable $x$ does not fully incorporate all active degrees of freedom, the right hand side in Eq.\ref{entropy} is only a lower bound of the true entropy production rate. Moreover, in practice one usually cannot estimate the full hierarchy of $m$-grams, so any partial result ($m<\infty$) again provides a lower bound to $\sigma_{\text{tot}}$. Also, when $x$ is defined on a continuous support, it is customary to symbolize it. In the case when $x$ is intrinsically discrete $x_i \in {\cal V}=\{v_1,\dots,v_{|{\cal V}|}\}$ then a proper estimate of the rhs in Eq.\ref{entropy} can only be attained when the sequence size $N\gg |{\cal V}|$ (exponentially larger): this is needed for collecting sufficient statistics on each $m$-gram. As we will show below, this latter observation is crucial in our context, as the musical compositions we consider here are relatively short symbolic sequences ($N$ ranges from a few hundreds to a few thousand points), while the number of symbols (notes) present in each musical piece is typically of the same order of magnitude as $N$, thus making the direct empirical estimation of eq.\ref{entropy} ineffective.\\
In what follows we introduce an alternative, graph-theoretic method  \cite{EPJB2012} which we show circumvents these issues.

\subsection{HVG-irreversibility}
A time series of $N$ points can be transformed into a so-called {\it horizontal visibility graph} (HVG) of $N$ nodes via the so-called horizontal visibility algorithm \cite{PNAS, PRE}. This is a non-parametric method that enables the characterisation of time series and their underlying dynamics using combinatorics and graph theory.

\begin{definition}
\label{HVG}
 Let ${\cal S}=\{x_1,\dots,x_N\}$, $x_i \in \mathbb{R}$ be a real-valued scalar sequence of $N$ data. Its horizontal visibility graph HVG(${\cal S}$) is defined as an undirected graph of $N$ vertices, where each vertex $i\in \{1,2,\dots,N\}$ is labelled in correspondence with the ordered datum $x_i$. Hence $x_1$ is related to vertex $i=1$, $x_2$ to vertex $i=2$, and so on.
Then, two vertices $i$, $j$ (assume $i<j$ without loss of generality) share an edge if and only if $x_k<\inf(x_i,x_j),\ \forall k: i<k<j$.
\end{definition}

\begin{figure}
\centering
\includegraphics[width=0.55\columnwidth]{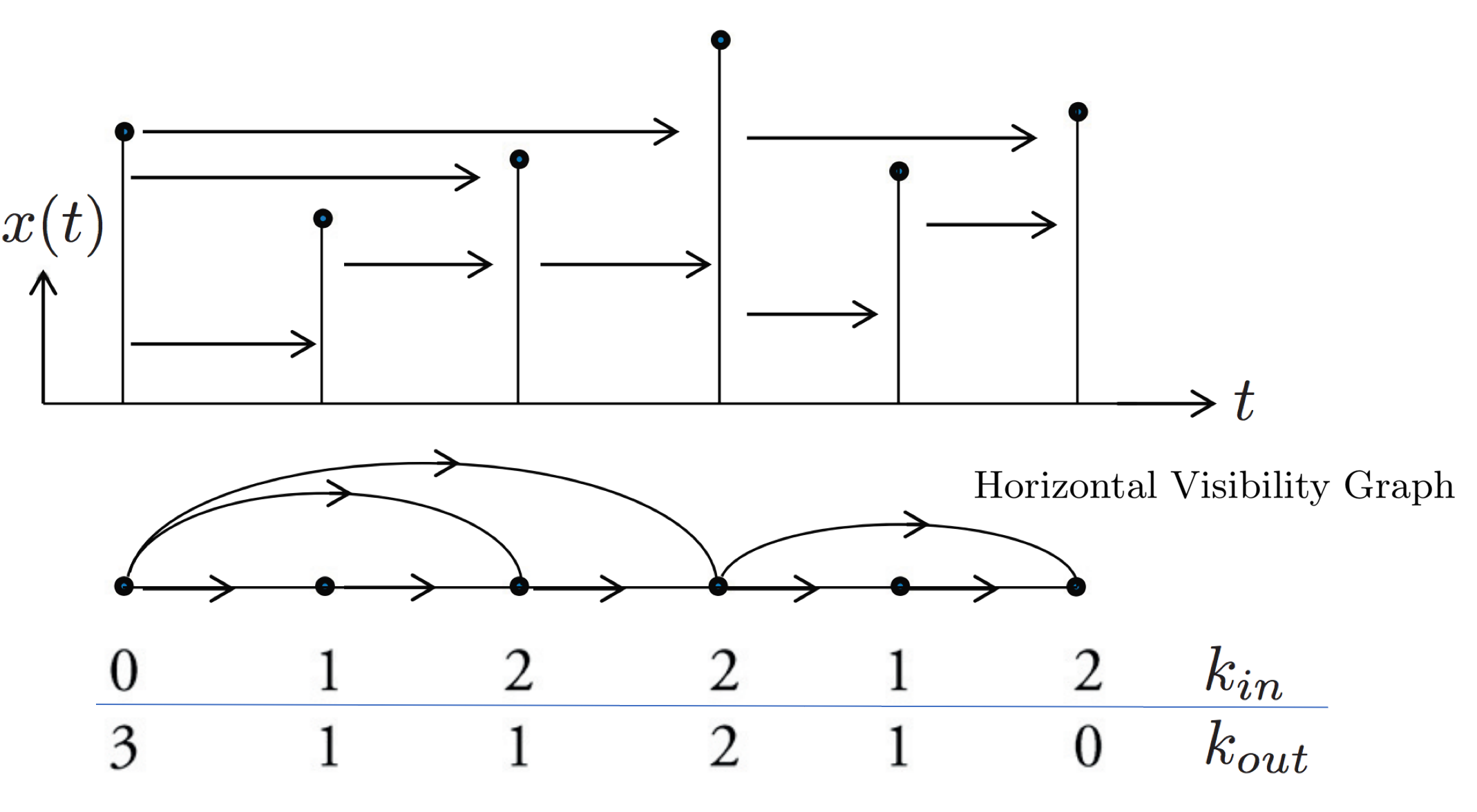}
\caption{Sample time series of $N=6$ data and its associated Horizontal Visibility Graph (HVG) (see \cite{EPJB2012} for details). By assigning a temporal arrow to each of the links, the degree sequence splits into an {\it in} degree and an {\it out} degree sequence, such that flipping the time series (time reversal operation) is equivalent to interchanging the in and out degree sequences. Assessing time reversibility in the time series reduces in this context to spot differences in the statistics of the in and out degree sequences, such as comparing the in and out degree distributions.}
\label{fig:HVG}
\end{figure}

\noindent
HVG implements an ordering criterion which can be visualized in Figure \ref{fig:HVG} (see \cite{PNAS} for a convexity criterion that generates `natural' visibility graphs instead).
Visibility and Horizontal Visibility graphs were introduced in the context of time series analysis with the aims of using the tools of Graph Theory and Network Science \cite{Newmanbook} to describe the structure of time series and their underlying dynamics from a combinatorial perspective (for other proposals for graph-theoretical time series analysis, see \cite{Kurths2017,PhysRep2018}).\\

\noindent
Among others, the concept of time series irreversibility has been recently explored within the context of visibility graphs \cite{EPJB2012,PRE2015,PLA2016}. In a nutshell,
if the HVGs of ${\cal S}$ and ${\cal S^*}$ have the same properties, then ${\cal S}$ is said to be HVG-reversible, and the concept has been shown to be applicable both in stationary and non-stationary processes  \cite{PRE2015}.
How is HVG-reversibility checked in practice? Since each node $i=1,\dots,N$ in the HVG is associated to a datum $x_i,\ i=1,\dots,N$ in the graph, there is a natural node ordering associated to the arrow of time. Such ordering is therefore inherited by the degree sequence, which has a natural representation ${\bf k}=(k_1,\dots,k_N)$, where $k_i$ is the degree of node $i$ (i.e., the number of links adjacent to node $i$). Now, while the HVG is initially an undirected graph, it can be converted into a directed one
by assigning a direction to each link in the HVG such that if $i<j$, then the link is $i \to j$. Assigning a direction to each of the links splits the degree sequence ${\bf k}={\bf k}_{\text{in}} + {\bf k}_{\text{out}}$, where ${\bf k}_{\text{in}}=(k^{\text{in}}_1,k^{\text{in}}_2,\dots,k^{\text{in}}_N)$ is the in-degree sequence and $k_i^{\text{in}}$ counts the number of links which are incident to node $i$, and respectively ${\bf k}_{\text{out}}=(k^{\text{out}}_1,k^{\text{out}}_2,\dots,k^{\text{out}}_N)$ where $k_i^{\text{out}}$ counts the number of links that emanate from node $i$.
Importantly, by construction one then has that the in-degree sequence of the HVG extracted from a given sequence is equal to the out-degree sequence of the HVG extracted from the time-reversed sequence, and therefore in order to assess time reversibility in ${\cal S}$, one can simply explore the statistical differences between the in-degree sequence and the out-degree sequence in the HVG (see \cite{EPJB2012,PRE2015} for details and Fig. \ref{fig:HVG} for an illustration).\\

\noindent In \cite{EPJB2012} some of us proposed that the right hand side Eq.\ref{entropy} could actually be approximated by comparing the {\it in} and {\it out} {\it degree distributions} of HVG(${\cal S}$), as these are the $(m=1)$-point marginal distribution of the {\it in} and {\it out} degree sequences. Note that only needing to look at $m=1$ statistics is a substantial difference with respect to the benchmark method based on comparing $m$-gram statistics of the time series, as in this latter case by construction the statistics of 1-grams are invariant under time reversal, and irreversibility can only be checked for $m=2$ or higher. This resulting reduction will enhance our ability to effectively use the method in short sequences, as we will show below.\\
\noindent Whereas originally the HVG method only looked at strings of size $m=1$ (1-grams) in the degree sequence, one could of course further generalise this measure if it were needed to account for blocks of arbitrary size in the degree sequence, following the spirit of Eq.\ref{entropy}.
For a block size $m$, let us consider strings of size $m$ within the {\it in} and {\it out} degree sequences such that with a little abuse of notation, ${\bf k}_{\text{in}}^m=(k^{\text{in}}_1,k^{\text{in}}_2,\dots,k^{\text{in}}_m)$ and similarly for ${\bf k}^m_{\text{out}}$. Additionally, let $P^{\text{in}}_m({\bf k})$ the marginal distribution of the in-degree sequence blocks ${\bf k}_{\text{in}}^m$  (respectively for $P^{\text{out}}_m({\bf k})$). Then, we now define the HVG-irreversibility of order $m$ as the Kullback-Leibler divergence between the in and out degree sequence blocks marginals
\begin{equation}
\text{KLD}_m(\text{in}||\text{out}) = \sum_{{\bf k}} P^{\text{in}}_m({\bf k}) \log \frac{P^{\text{in}}_m({\bf k})}{P^{\text{out}}_m({\bf k})},
\label{kld}
\end{equation}
This quantity is null if and only if the size-$m$ blocks are equally distributed in the (infinitely long) degree sequence, and is positive otherwise.\\
To summarise, the procedure to evaluate HVG-reversibility in a time series ${\cal S}$ is as follows:
\begin{itemize}
\item From ${\cal S}$ we construct the directed HVG, and subsequently extract the {\it in} and {\it out} degree sequences ${\bf k}_{\text{in}}$ and ${\bf k}_{\text{out}}$. Note that for ${\cal S^*}$,  ${\bf k}_{\text{in}}$ and ${\bf k}_{\text{out}}$ are interchanged. This procedure is applicable when ${\cal S}$ comes from both stationary and non-stationary processes alike.
\item To quantify for HVG-reversibility (and HVG-entropy production), we make use of Eq. \ref{kld}.
Note that $m=1$ is the smallest nontrivial case here, as 1-grams of the degree sequence already incorporate temporal directionality (when using $m$-grams of $\cal S$ vs $\cal S^*$, $m=2$ is the simplest nontrivial case instead).
\end{itemize}

\subsection{On the number of symbols}
\begin{figure}
\centering
\includegraphics[width=0.4\columnwidth]{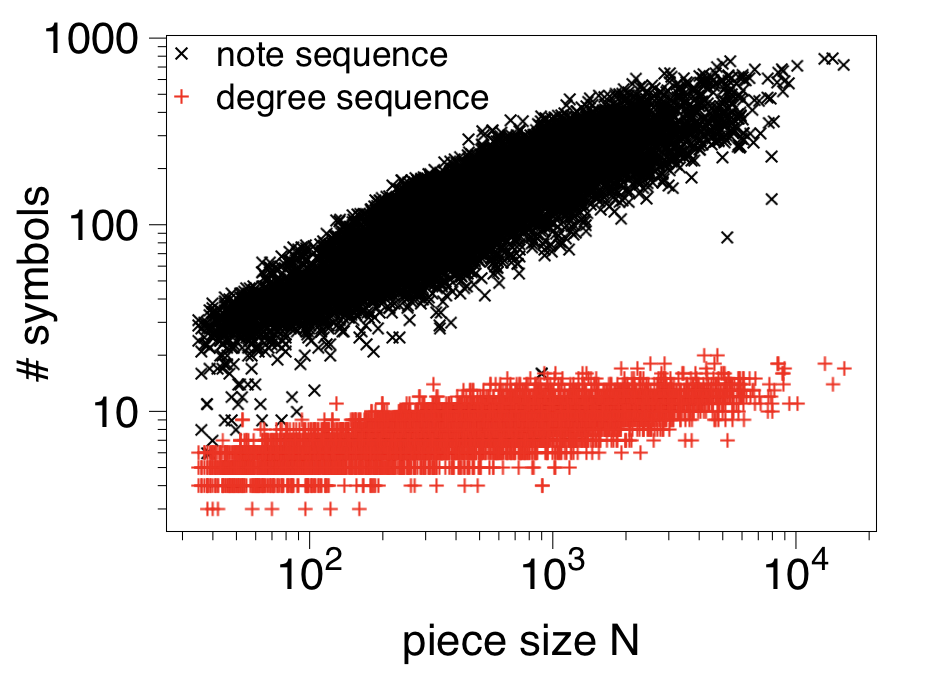}
\includegraphics[width=0.4\columnwidth]{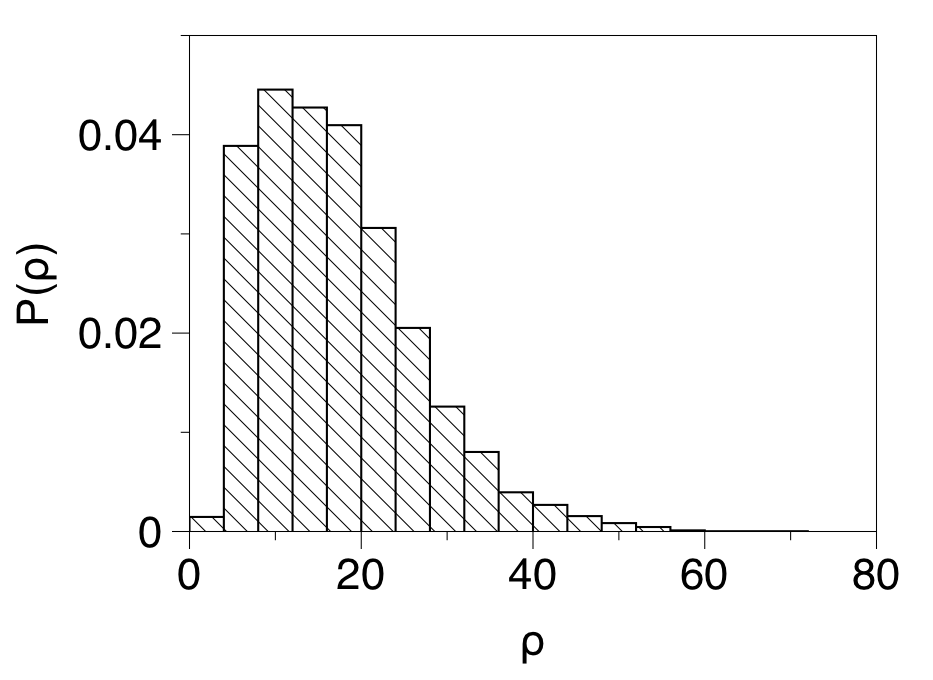}
\caption{(Left panel) Log-log plot of the minimum number of effective symbols needed to determine time irreversibility of a given musical composition, as a function of the size $N$ of the composition, using our HVG-based method (red plus) and using a standard $2$-gram comparison in the note sequence (black crosses). We can see that the number of symbols in the method based on the note sequence is systematically in the same order of magnitude of total size of the sequence, hence finite size effects are too strong and make the method unapplicable. The method based on the HVG requires a substantially smaller number of symbols, orders of magnitude smaller than the sequence size, hence making this method useful in this dataset.
(Right panel) Histogram of the reduction factor $\rho$ (see the text) which compares the effective number of symbols needed in the HVG irreversibility method with respect to a standard method based on $n$-gram statistics. When we average over all the musical pieces, the average number of symbols is reduced 17 times, and this reduction can reach up to 75 times in some cases. This strongly reduces the finite size effects and hence makes the
HVG method  useful to explore time irreversibility in short sequences.}
\label{fig:reduction}
\end{figure}
Once we have outlined the procedure to estimate the HVG-irreversibility, we now consider the problem of working with short experimental sequences.
First, it is important to note that the {\it in} and {\it out} degrees typically take values from a {\it small} alphabet --systematically smaller than the original musical note alphabet as we will show below--, whose size increase at most logarithmically with $N$. This is because the probability that an arbitrary node in a HVG has a certain {\it in} and {\it out} degree $k$ typically decays exponentially fast with $k$ \cite{nonlinearity}. Moreoever, it is important to recall that temporal irreversibility can already be assessed for $m=1$, that is, by only looking at the marginal distribution of the {\it in} and {\it out} degree sequences. Conversely, if we were to estimate time irreversibility directly on the note sequences, then we would at least need to consider strings of size $m=2$ consecutive notes. The effective number of symbols needed is therefore much larger than in our HVG setting. In the left panel of figure \ref{fig:reduction} we plot the effective number of symbols required to assess time irreversibility using HVG (red plus sign) and using a simple $2$-gram comparison of the forward and backward note sequence (black crosses). We see that in most of the cases, the number of note 2-grams is of the same order of magnitude than the size of the note sequence, hence making the standard approach useless. On the other hand, the approach based on HVG keeps the number of symbols needed to a bare minimum and is therefore useful in this context.\\
To give an additional quantitative idea of the symbol reduction given by the HVG method, let us define a reduction factor $\rho$ associated to a given musical piece as
$$\rho=\frac{\text{total number of strings of size $m=2$ empirically found in the MIDI note sequence}}{\text{total number of different (in our out) degrees empirically found in the in and out degree sequence}}.$$
In figure \ref{fig:reduction} we plot the histogram $P(\rho)$ estimated over our whole dataset. The average reduction is about 17, which means that our irreversibility method yields on average a $1700 \%$ reduction over the standard method based on counting the statistics directly on the note sequence.
This implies that the finite size effects (subsampling) --which will appear due to the fact that the musical composition are not exponentially larger than the number of symbols-- will be contained in the case of the HVG method, thus enabling its use in applications where time series are short, such as in musical compositions.\\

\noindent Having said that, while small, finite size effects are still expected to emerge in finite time series, and because of that, the Kullback-Leibler divergence will always be positive (vanishing for reversible processes only asymptotically), which poses some interpretability problems. In what follows we introduce a confidence index whose aim is to solve this interpretability issue.\\

\subsection{Irreversibility ratio $\text{IR}_m$: a confidence index}
Let us define the $m$-order Irreversibility Ratio $\text{IR}_m$ by standardizing the net irreversibility measure $\text{KLD}_m(\text{in}||\text{out})$ with respect to a null model where the original sequences are shuffled:
\begin{equation}
\text{IR}_m = \frac{ \text{KLD}_m(\text{in}||\text{out})  - \langle\text{KLD}_m(\text{in}||\text{out}) \rangle_{\text{null model}}}{\sigma[\text{KLD}_m(\text{in}||\text{out})]_{\text{null model}}}
\label{IR}
\end{equation}
where $\sigma$ is the standard deviation.

\noindent The process of standardizing is applied in order to be able to compare results across samples (musical pieces) with different sizes and marginal distributions, something which is recurrent in the musical compositions we analyse. More concretely, in the ideal situation of estimating irreversibility for infinitely long sequences, $\text{KLD}_m(\text{in}||\text{out})$ vanishes if and only if $P^{\text{in}}_m({\bf k}) = P^{\text{out}}_m({\bf k})$, and therefore values different from zero would indicate that the sequence is statistically HVG-irreversible. However, in practice this is not so clear cut:
the quantity $\text{KLD}_m(\text{in}||\text{out})$ for a reversible process is only asymptotically null (vanishing as the sample size goes to infinity), and will be finite (yet small) for finite samples due to statistical deviations in the estimation of joint probability functions $P({\bf k}_{\text{in}})$ and $P({\bf k}_{\text{out}})$. By standardizing this quantity with respect to a suitable null model, one can quantify the effective deviation of a given finite sample from the expected value if that finite sample was generated by a truly reversible process. The null model is built by taking 200 randomizations of the sample sequence, and computing $\text{KLD}_m(\text{in}||\text{out})$ on each randomized sample. Then Eq.\ref{IR} measures the effective distance of a sample to its null model, in standard deviation (aka `sigma') units. For instance, if $\text{IR}_m\leq 1$ this means that the irreversibility value of a sample finite sequence is not statistically distinguishable from a finite sample of the same size and marginal distribution extracted from a truly reversible process. That does not necessarily mean that the process is HVG-reversible, it only means that there is no statistical significance to assert otherwise.
Similarly, if e.g. $\text{IR}_m = 2$, this means that the HVG-irreversibility value of the sample is two standard deviations larger than the one expected for its (reversible) null model.\\
\begin{figure}[htb]
\centering
\includegraphics[width=0.65\columnwidth]{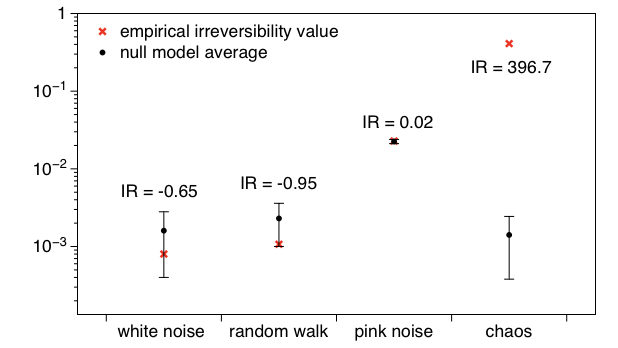}
\caption{Irreversibility ratios IR$_1$ for three theoretically HVG-reversible processes (uniform white noise, unbiased random walk, pink noise) and one theoretically HVG-irreversible process (fully chaotic logistic map $x_{n+1} = 4 x_n(1-x_n)$). In every case we generate a time series of $N=10^4$ data points and symbolize with a vocabulary of $|{\cal V}|=100$ symbols. Red crosses correspond to the empirical irreversibility value of order 1 of the time series $\text{KLD}_1(\text{in}||\text{out}) $, black dots correspond to the ensemble average irreversibility value  $\langle\text{KLD}_1(\text{in}||\text{out}) \rangle_{\text{null model}}$ of $10^3$ null models constructed by randomizing the empirical time series (see the text), and the brackets correspond to $\pm$ one standard deviation (note that the Y axis is in logarithmic scales). The irreversibility ratio of order 1 (see the text) decides whether the empirical time series is reversible (IR$_1\leq1$) or irreversible (IR$_1>1$), and correctly predicts the true nature of the process. Interestingly, the empirical irreversibility value $\text{KLD}_1(\text{in}||\text{out})$ of pink noise is notably larger than for white noise or random walk, which could mislead to the suggestion that pink noise is irreversible. The irreversibility ratio asserts that this is not the case, suggesting that the difference in the raw values is due to differences in the specific marginal distribution of the underlying processes, which only yield spurious effects on the determination of temporal irreversibility.}
\label{fig:IR}
\end{figure}

\noindent For illustration and validation, in Figure \ref{fig:IR} we plot the raw HVG-irreversibility value $\text{KLD}_1(\text{in}||\text{out})$ (red crosses) and IR$_1$ for time series extracted from four synthetic dynamical processes: (i) white noise, (ii) random walk, (iii) pink noise and (iv) a fully chaotic process.\\
Process (i) is an (uncorrelated) uniform white noise. This is a stationary stochastic process with delta-like autocorrelation function, it is statistically reversible and HVG-reversible. The method correctly identifies this character as IR$_1<1$ in this case.\\
Process (ii) is an unbiased discrete random walk $x(t+1)=x(t)+\eta$, where $\eta \sim \text{uniform}(-1/2,1/2)$. Interestingly, while this is a non-stationary process and thus could be seen as irreversible, it can be shown to be so-called HVG-stationary \cite{PRE2015} and explored adequately within the HVG framework, as HVG-reversible with IR$_1<1$. This is indeed convenient as Brownian particles do not produce entropy on average, so in the context of HVG-reversibility, we recover the relation between reversibility and entropy production in this non-stationary process.\\
Process (iii) is a {\it linearly} correlated noise with a $1/f$ spectrum. This is again, by definition, a time reversible process, and is a paradigmatic (stationary) stochastic process to describe music from the pioneering works of Voss and Clarke \cite{Voss1, Voss2}. Also in this case we correctly detect the reversible character as IR$_1<1$.\\
Finally, process (iv) is a deterministic chaotic process generated by a fully chaotic logistic map $x_{t+1}=4x_t(1-x_t)$. This process is dissipative and statistically time irreversible (and HVG-irreversible), as certified by IR$_1>1$.\\

\begin{figure}
\centering
\includegraphics[width=0.45\columnwidth]{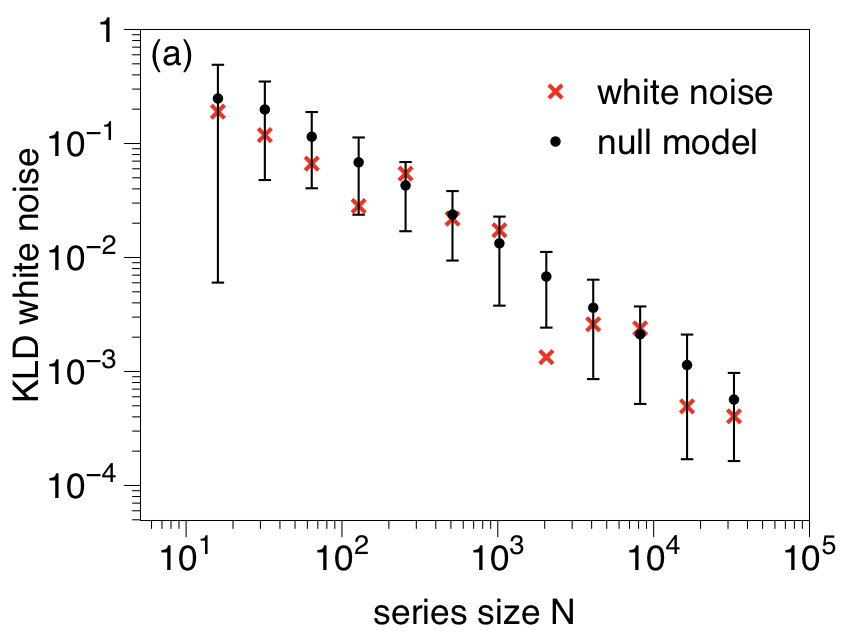}
\includegraphics[width=0.45\columnwidth]{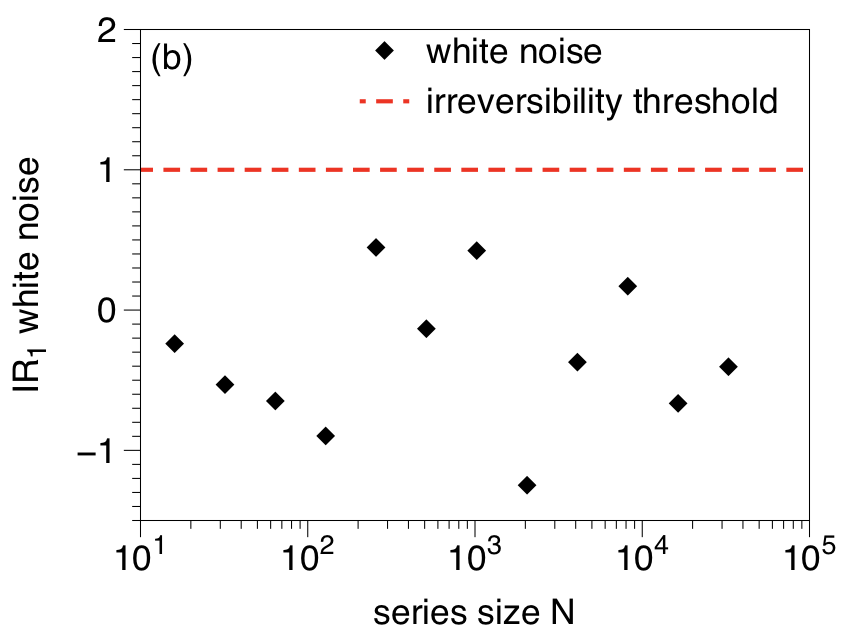}
\includegraphics[width=0.45\columnwidth]{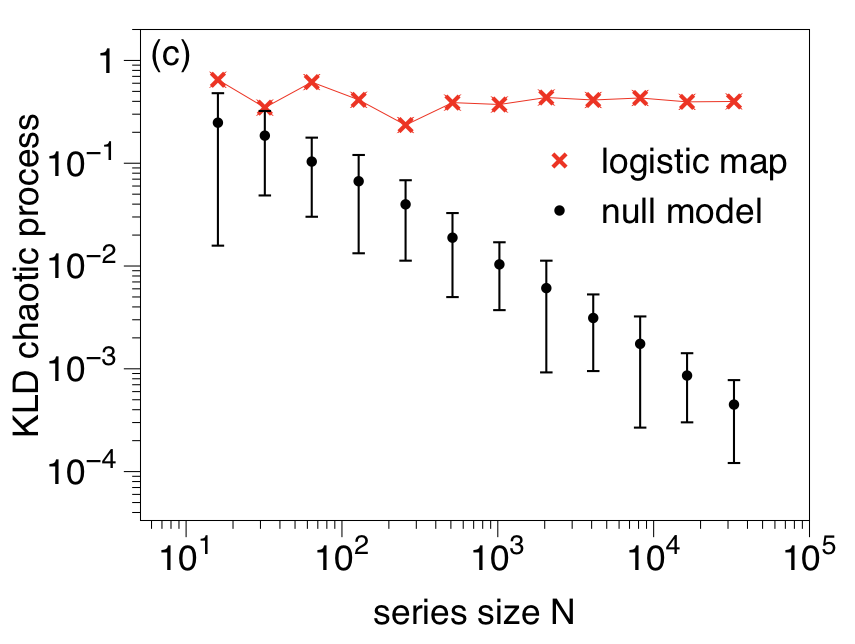}
\includegraphics[width=0.45\columnwidth]{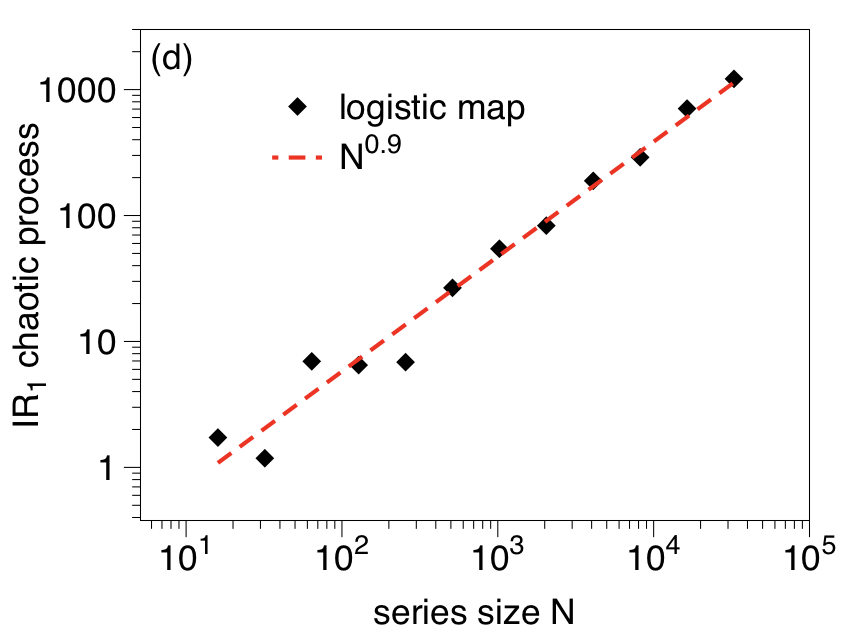}
\caption{({\bf a}) Raw order-1 irreversibility value $\text{KLD}_1(\text{in}||\text{out})$ of a time series of size $N$ generated by a reversible white noise process $x_{t}=\eta, \ \eta\sim \text{Uniform(0,1)}$ (red crosses). For comparison, the results of a null model (mean $\pm$ one standard deviation) where we randomize the original time series 200 times and compute the irreversibility value is shown in black solid circles. The raw irreversibility value decreases without bounds in a fashion similar to the null model, certifying that positive irreversibility values are here only due to finite size effects which vanish as $N$ increases. ({\bf b}) Order-1 irreversibility ratio IR$_1$ for the process depicted in panel (a), certifying that the time series is not distinguishable from a reversible process for any time series size.
({\bf c}) Similar to (a),  for a time series of size $N$ generated by an irreversible, fully chaotic logistic map $x_{t+1}=4x_t(1-x_t)$ (red crosses). The raw irreversibility value is always positive and stabilizes for $N>10^3$. The null model is by definition reversible, and its irreversibility value is only positive due to finite size effects, hence decreases as $N$ increases. Distinguishability therefore increases with $N$. ({\bf d}) Similar to (b), for the chaotic case, certifying that the process is irreversible and that the level of confidence increases without bounds.}
\label{fig:LOG}
\end{figure}

\noindent Another important aspect is to understand how IR$_m$ is affected by the time series length $N$. Intuitively, if the underlying process is HVG-reversible, then $\text{KLD}_1(\text{in}||\text{out})$ should be similar to its null model (within its uncertainty) and both measures should be decaying with the same trend as $N$ increases, and therefore one should expect IR$_m<1$, independently of $N$. If, on the other hand, the underlying process is HVG-irreversible, then $\text{KLD}_1(\text{in}||\text{out})$ should be systematically larger than zero and remain positive as we increase $N$. Since the null model of this process is HVG-reversible, its irreversibility value should decrease as $N$ is increased, we thus expect IR$_m$ effectively to increase with $N$ without bounds. This only means that for an HVG-irreversible process, whereas for short time series the distinguishability is small (IR$_m$ close to 1), as we increase the series size it is systematically easier to ascertain that the series (and process) is HVG-irreversible. In some sense, IR$_m$ is therefore a {\it measure of confidence}.\\
We illustrate this dependence on series size in Fig \ref{fig:LOG}, where in panels (a,b) we consider a truly reversible process (white noise $x_t=\eta, \eta \sim \text{Uniform}(0,1)$) and in panels (c,d) we consider a truly irreversible process (a fully chaotic logistic map $x_{t+1}=4x_t(1-x_t)$). Results indeed confirm our discussion above.\\

\noindent Once we have illustrated how the irreversibility ratio works, let us introduce a classification of net HVG-irreversibility for a given musical piece as it follows:

\begin{definition}
\label{classes}
A musical piece is defined as HVG-reversible (or simply reversible) if $\text{IR}_1\leq1$. If $1<\text{IR}_1\leq4$ we say that the process is HVG-irreversible with {\it weak confidence}. If $4<\text{IR}_1\leq10$, we say that it is HVG-irreversible with {\it strong confidence}, and if  $\text{IR}_1>10$ we say that the musical piece is HVG-irreversible with {\it extreme confidence}.
\end{definition}

\noindent A similar classification can be defined for higher orders $m>1$, however this is not needed for this work since $\text{IR}_1$ shows strong correlation with $\text{IR}_{m>1}$ for most of the musical pieces considered, hence it will be enough to concentrate our analysis on $m=1$. Incidentally, note that since blocks are extracted from the HVG's degree sequence (not directly from the sequence of notes), in principle the HVG-irreversibility measure at $m=1$ already gathers temporal information of different time scales.\\

\noindent Summing up, we have shown that the concept of HVG-reversibility is better suited than the standard concept of statistical time reversibility to investigate the arrow of time in stationary and nonstationary processes \cite{PRE2015}, and HVG is a useful approach when time series are short.
We can safely conclude that IR$_m$ is a measure that quantifies our certainty that the time series under study was generated by an HVG-irreversible process. To compute such confidence, the size of the series (and the resulting finite-size effects) and the specific shape of the marginal distribution of the signal must be taken into account in order to provide a quantifier which is not affected by these variables. On the other hand, and once the analysis based on IR$_m$ leads us to conclude that the process is indeed HVG-irreversible, we will then use $\text{KLD}_m(\text{in}||\text{out})$  as a bound for the true (thermodynamic) entropy production rate of the process.\\
In the next sections we will  compute HVG-reversibility metrics, but in order to lower down verbosity we will refer to it indistinctively as either HVG-reversibility or just reversibility.

\section{Experimental Results}



\subsection{Irreversibility}
\begin{figure*}
\centering
\includegraphics[width=0.3\columnwidth]{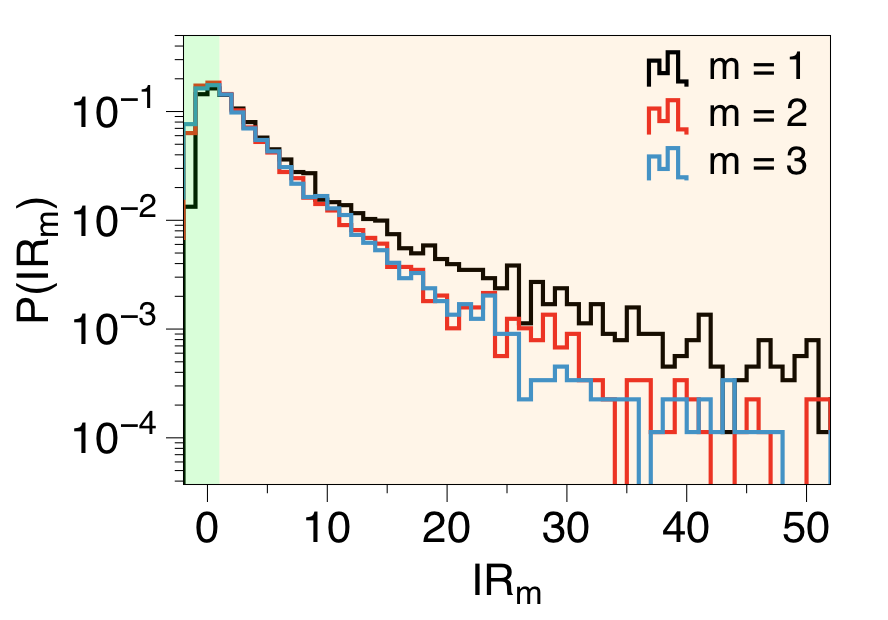}
\includegraphics[width=0.3\columnwidth]{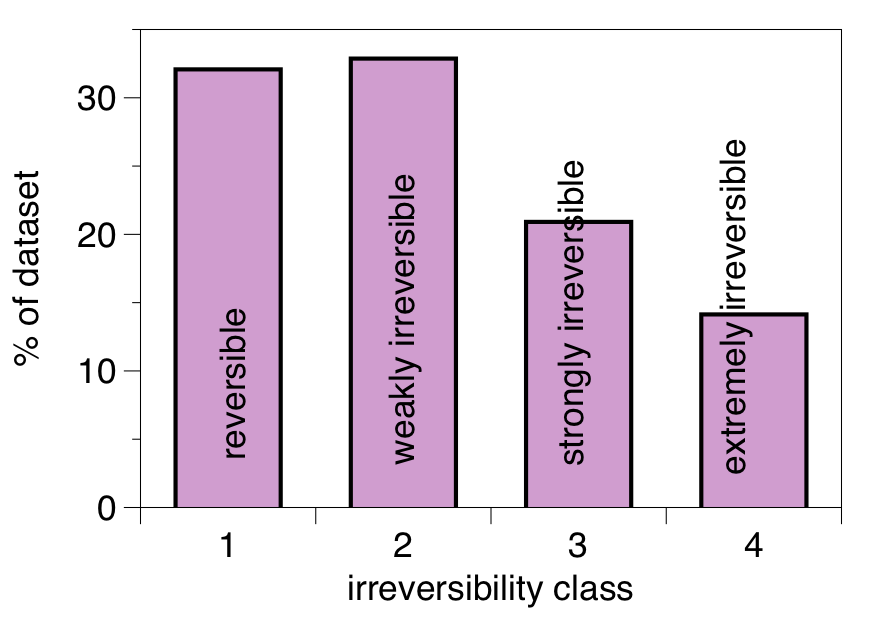}
\includegraphics[width=0.3\columnwidth]{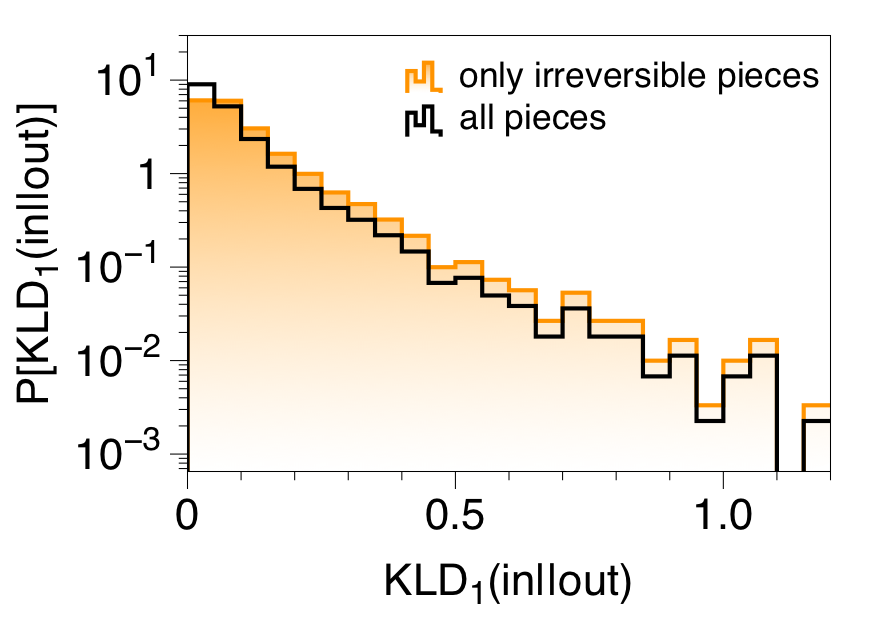}
\caption{(Left) Estimated normalized histogram of irreversibility ratios IR$_m$ ($m=1,2,3$) for all pieces in our dataset, in a semi-log plot. The green area highlights the region where pieces are reversible. A large portion of the pieces are indeed generated by an irreversible process.
 (Middle) Abundance per irreversibility confidence class. (Right) Estimated normalized histogram of raw irreversibility values $\text{KLD}_1(\text{in}||\text{out})$, for all pieces (black) and only those pieces which have previously been certified as irreversible (i.e. IR$_1>1$) (orange). A linear binning of size $0.05$ as been performed.}
\label{fig:histos}
\end{figure*}
To start, we have explored the confidence of time series irreversibility --as quantified by IR$_m$-- for all the pieces considered in this work. In the left panel of figure \ref{fig:histos} we depict in semi-log scales the normalised histograms $P(\text{IR}_m)$ for $m=1,2,3$. The orange area denotes the region IR$_m>1$, showing that a large percentage of pieces are irreversible (HVG-irreversible) at all $m$ orders, with varying degree of confidence.\\
\noindent A second observation is that the histograms for $m=2$ and $m=3$ are indeed very similar. In order to further understand to which extent the three measures IR$_1$, IR$_2$ and IR$_3$ are correlated, we have computed the Pearson correlation coefficient between IR$_m$ and IR$_{m-1}$ (see appendix figure \ref{fig:scatter}), showing that indeed we find a strong correlation between all of them. The interpretation of this finding is two-fold: we can first argue that, in the context of classical music and the database analysed in this work, higher-order irreversibility is irrelevant, and all the structure can be efficiently captured by order-1 HVG-reversibility. Second, this also means that from now on we can safely focus our analysis on $m=1$, which is notably faster to estimate.\\
The abundance of each irreversibility confidence class is depicted in the middle panel of figure \ref{fig:histos}, showing that all four classes have a notable representation, where only about {30}\% of the whole dataset complies with a reversible structure.
In the right panel of the same figure we have plotted in semi-log the normalized histogram of $\text{KLD}_1(\text{in}||\text{out})$, for all pieces (black curve) and only for those pieces which have previously been certified to be irreversible (IR$_1>1$). In both cases, values mostly concentrate in the interval $[0,1]$, and most of the pieces with high irreversibility value (e.g. $\text{KLD}_1(\text{in}||\text{out})>0.1$) correspond to those pieces previously checked to be indeed irreversible according to the irreversibility ratio criterion. Distributions decay rapidly in both cases, highlighting that this measure is highly concentrated towards the left end of the spectrum.\\

\begin{table}[htp]
\begin{center}
\begin{tabular}{|c|c|c|c|c|}
\hline
{\bf Rank}&{\bf Composer}&$\langle\text{KLD}_1(\text{in}||\text{out})\rangle$&{\bf DOB}&{\bf number of pieces}\\
\hline
1&Bizet&0.28&1838&21\\
2&Mahler&0.19&1860&28\\
3&Dowland&0.18&1563&61\\
4&Desprez&0.178&1521&34\\
5&Paganini&0.155&1782&24\\
\hline
\end{tabular}
\end{center}
\vspace{-3mm}
\caption{Ranking of most irreversible composers.}
\label{table:composers}
\end{table}%

\noindent We can now concentrate on the subset of pieces which are certified to be irreversible (IR$_1>1$), and we can {\it rank} both composers and pieces according to their net irreversibility value. For composers, this ranking is shown in table \ref{table:composers}. Since irreversibility is linked with entropy production, we could provocatively say that this is a ranking of the composers which, on average, have a `more out of equilibrium' compositional process, i.e. the composers whose compositional style dissipates more energy and on average produce more entropy accordingly. A similar ranking for the most irreversible pieces is depicted in table \ref{table:pieces}.


\begin{table}[htp]
\begin{center}
\begin{tabular}{|c|c|c|c|c|c|}
\hline
{\bf Rank}&{\bf Piece}&$\text{KLD}_1(\text{in}||\text{out})$&{\bf Composer}&{\bf DOB}&{\bf Piece length}\\
\hline
1&Silhouettes 09&5.33&Dvorak&1841&269\\
2&Gadfly 97a,3&3.82&Shostakovitch&1906&492\\
3&Swan Lake (Act4,25 Entract e)&3.57&Tchaikovsky&1840&179\\
4&Kinderszenen 15 10&2.45&Schumann&1810&199\\
5&Messiah 14&2.22&Handel&1685&96\\
\hline
\end{tabular}
\end{center}
\vspace{-3mm}
\caption{Ranking of most irreversible pieces.}
\label{table:pieces}
\end{table}%

\begin{figure}[htb]
\centering
\includegraphics[width=0.31\columnwidth]{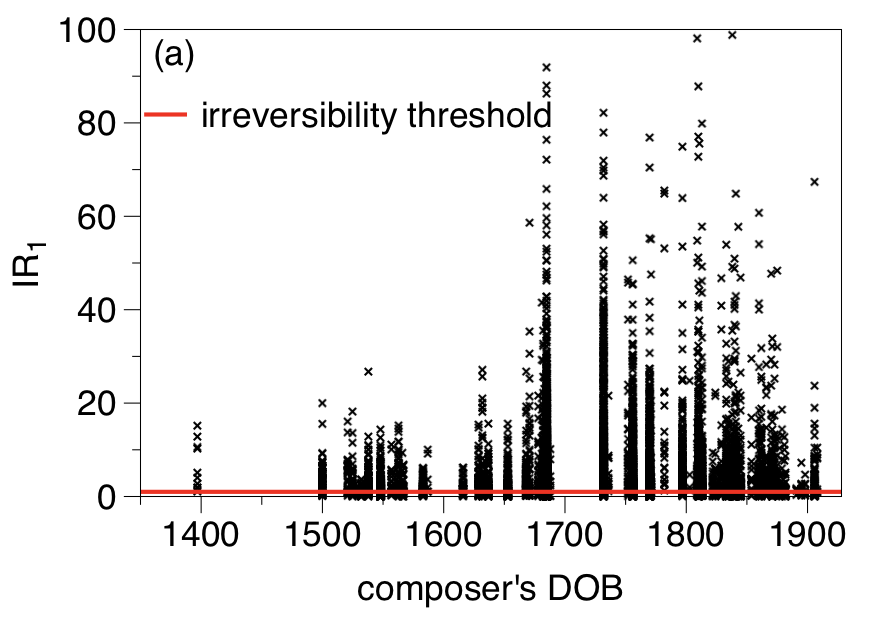}
\includegraphics[width=0.31\columnwidth]{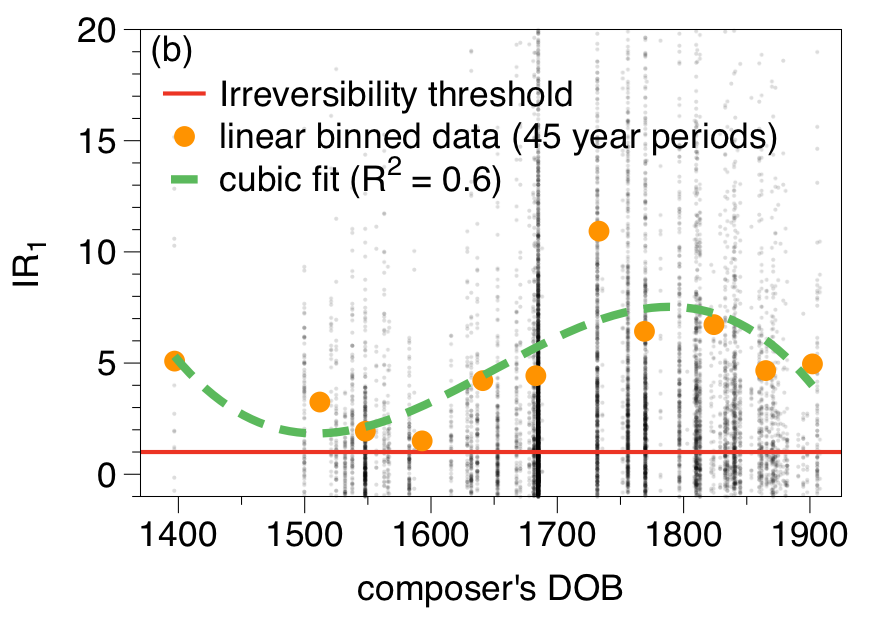}
\includegraphics[width=0.31\columnwidth]{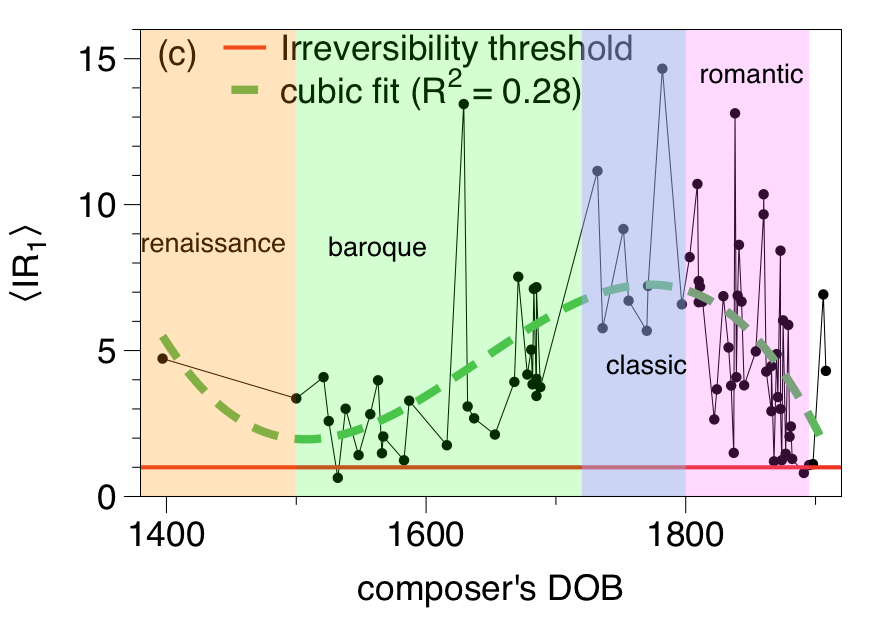}
\includegraphics[width=0.31\columnwidth]{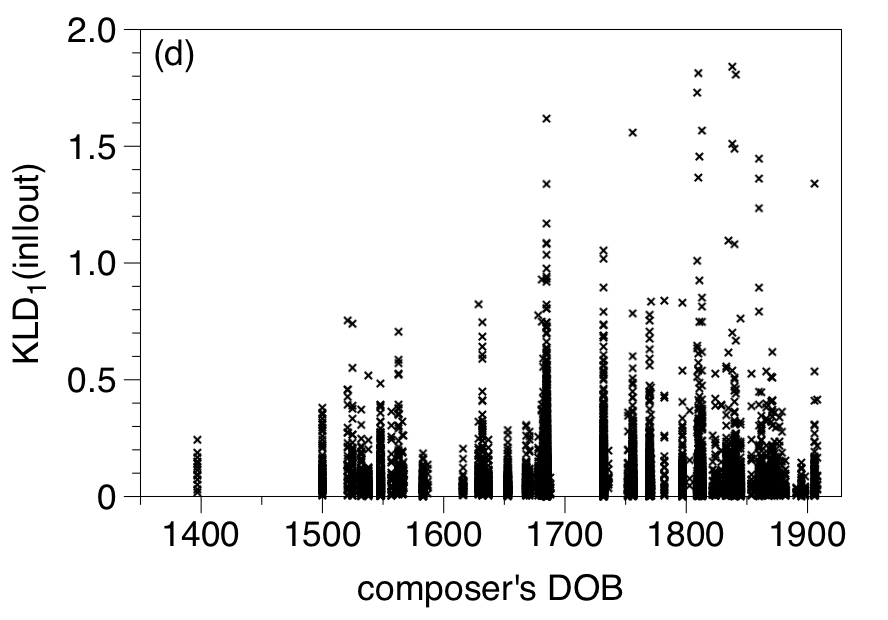}
\includegraphics[width=0.31\columnwidth]{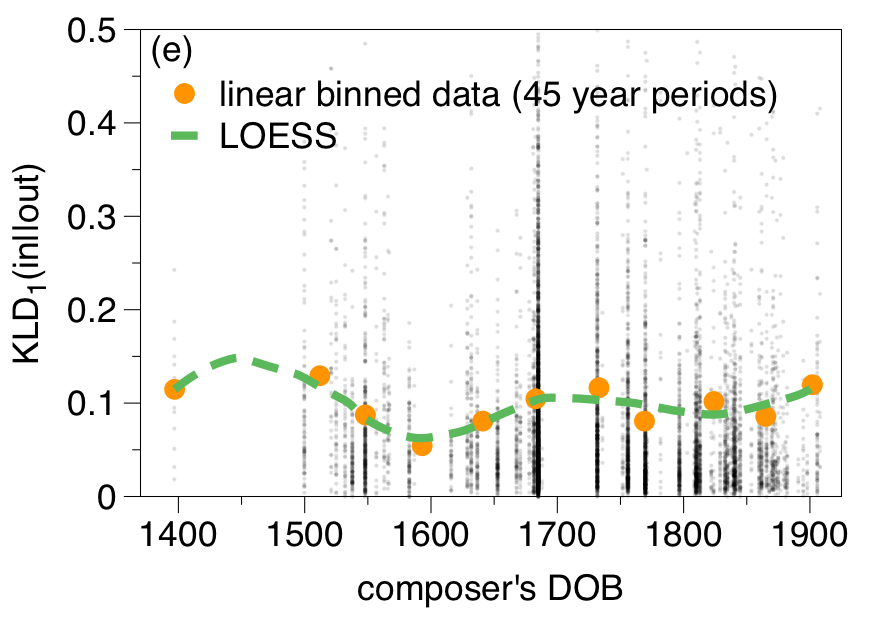}
\includegraphics[width=0.31\columnwidth]{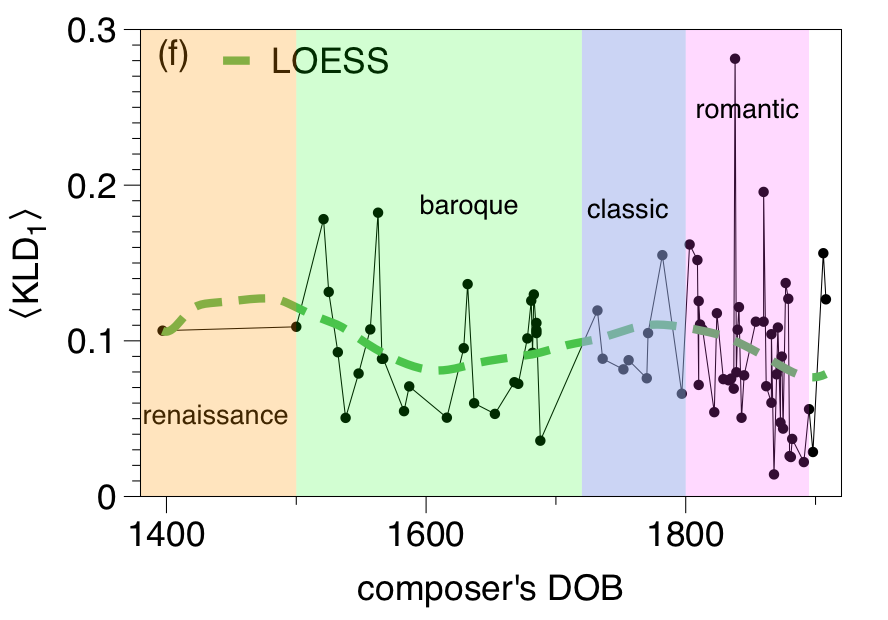}
\caption{{\bf Evolution of irreversibility over different musical periods. }({\bf a}) Irreversibility ratio IR$_1$ vs the composer's date of birth of each piece (crosses), for all pieces considered. The red solid line denotes the irreversibility threshold, suggesting that most of the pieces are statistically irreversible. ({\bf b}) Similar to panel (a), but where a cubic function (green dashed line) has been fitted to data after a linear binning (in periods of 45 years) has been performed. The fit highlights a modulating trend by which compositions net irreversibility confidence tend to vary over different periods. ({\bf c})  Similar to the (a) panel, but where irreversibility ratio is averaged over all the pieces of the same composer (one data point per composer). The modulating trend is still present, and for reference we also include the different musical periods (Renaissance, Baroque, Classical, Romantic and early Modern). ({\bf d}) Similar to (a), but where we plot the raw irreversibility value $\text{KLD}_1(\text{in}||\text{out})$ as a proxy for the entropy production rate per piece. Qualitatively the same pattern as in (a) emerges. ({\bf e}) Similar to (b) but where we plot $\text{KLD}_1(\text{in}||\text{out})$. Orange dots correspond to a linear binning and can be interpreted as the average entropy production rate of all pieces over each 45-year periods. This rate is maintained approximately constant over time. The fit is a local estimated scatterplot smoothing (LOESS). ({\bf f}) Similar to (c), but averaging the value $\text{KLD}_1(\text{in}||\text{out})$ for all pieces of each artist separately. This gives us a proxy for an average entropy production rate per composer. }
\label{fig:date}
\end{figure}
\noindent In order to explore the evolution of irreversibility over different periods, in Fig. \ref{fig:date} we plot the values of IR$_1$ (top) and $\text{KLD}_1(\text{in}||\text{out})$ (bottom) for all pieces as a function of the date of birth of the composer of the piece. Interestingly, the irreversibility ratios seem to fluctuate in a non-random way. To highlight such modulation, in panel (b) of figure \ref{fig:date} we plot the same data and we add as orange dots the result of a linear binning (bins of 45 year size). The green dashed dot corresponds to the best fit of a cubic polynomial to the binned data, highlighting an apparent modulation inside musical compositions over different time periods. In panel (c) of the same figure, instead of displaying the IR$_1$ for each piece, we average per composer and plot the average irreversibility ratio $\langle \text{IR}_1 \rangle$ for each composer, as a function of the composer's date of birth. A cubic polynomial is once more fitted to this data, again highlighted the modulated trend. Superposed to this figure we have highlighted the different musical periods (Renaissance, Baroque, Classic, Romantic and early Modern). Since IR$_1$ has a strong dependence on piece size and this is not the case for $\text{KLD}_1(\text{in}||\text{out})$ (see appendix C), a similar analysis is then replicated in panels ($d-f$) of Fig.\ref{fig:date} for the irreversibility value $\text{KLD}_1(\text{in}||\text{out})$, tracking how irreversibility values (i.e., entropy production) evolves over time. The modulated trend along the different musical periods is not anymore present. We can only conclude that over different periods, different composers have systematically produced pieces which were irreversible, but the effect of different styles in the particular value of irreversibility is less clear.

\begin{figure}[htb]
\centering
\includegraphics[width=0.35\columnwidth]{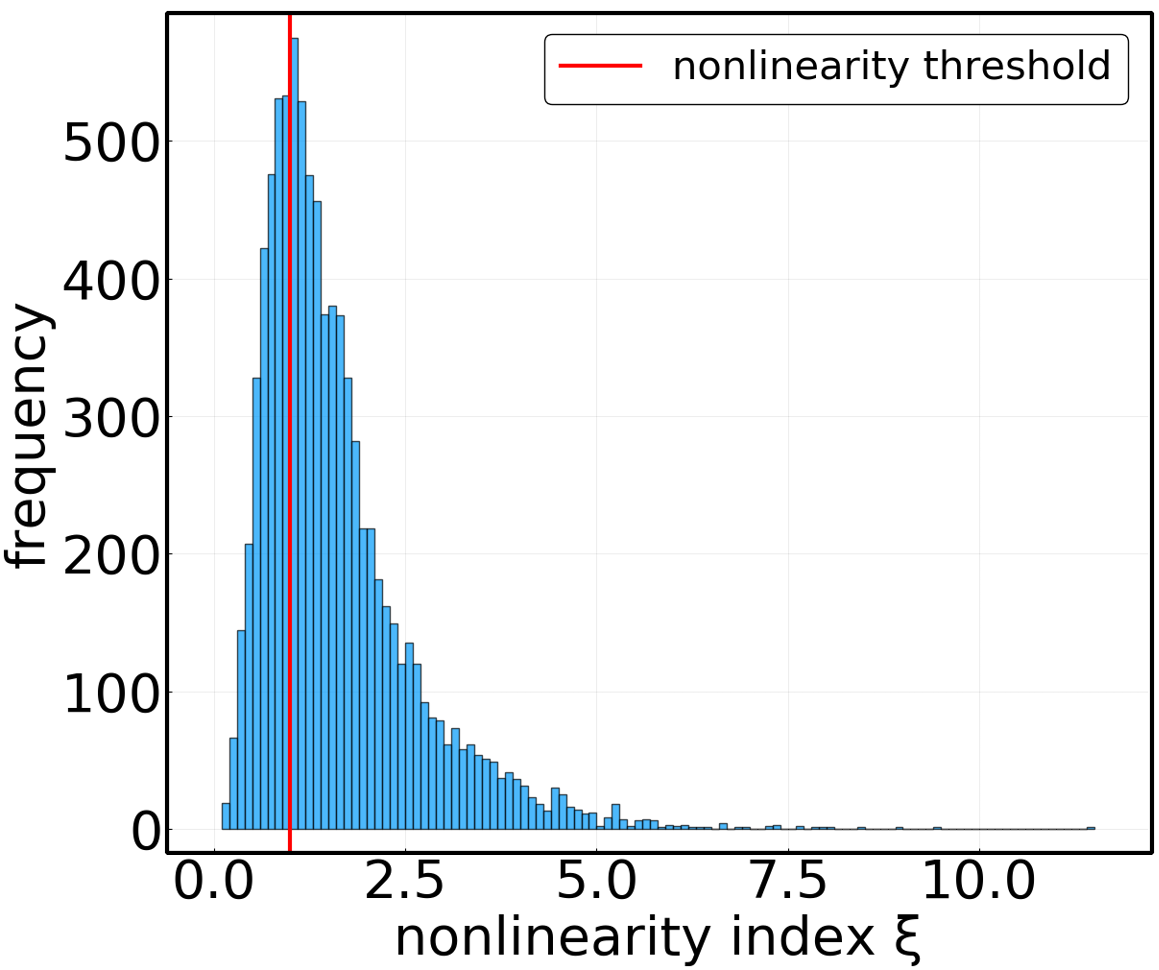}
\includegraphics[width=0.35\columnwidth]{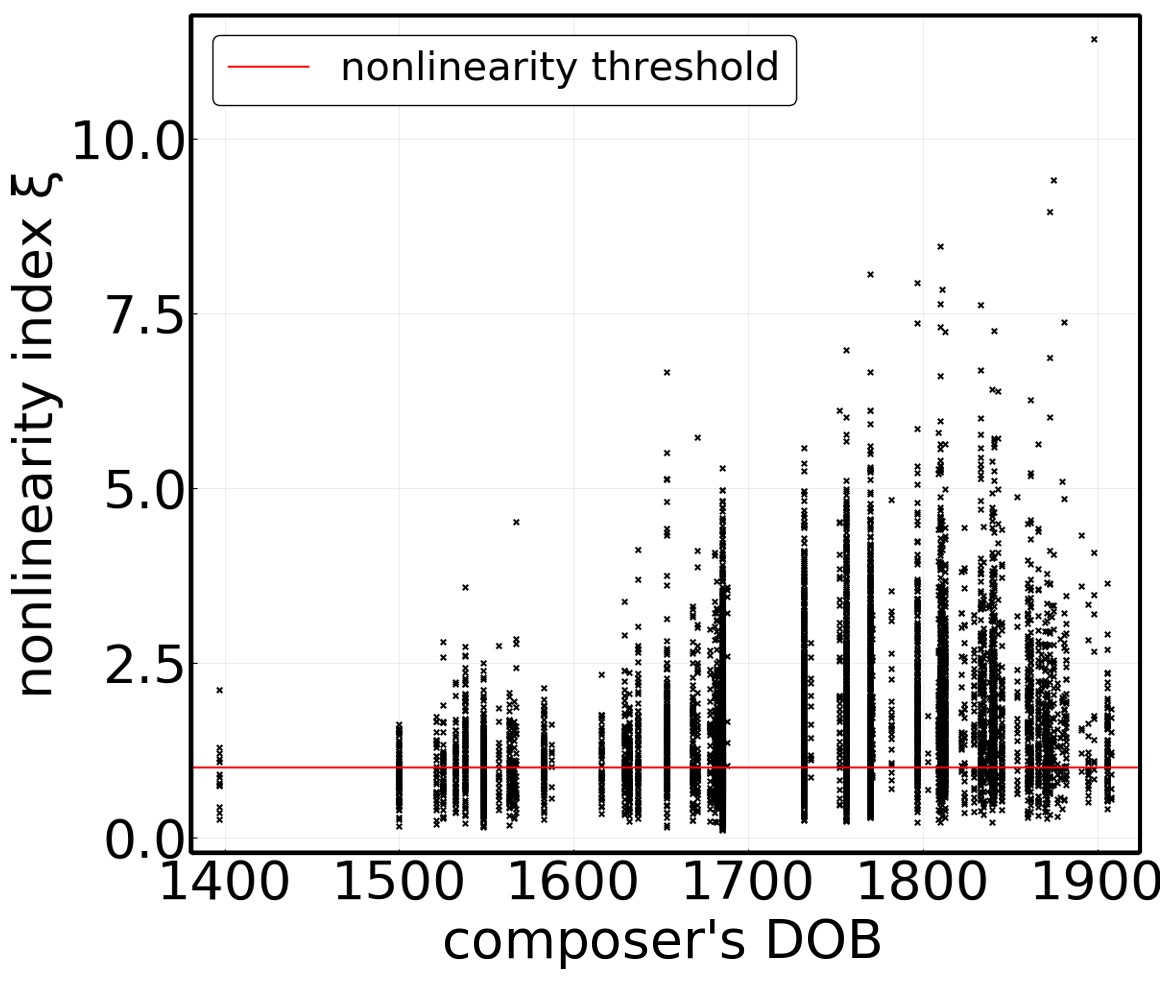}
\caption{{\bf Nonlinearity index. }(Left panel) Histogram of the nonlinearity index $\xi$ (see the text) for all pieces considered. (Right panel) Nonlinearity index $\xi$ as a function of the composer's date of birth.}
\label{fig:nonlin}
\end{figure}

\subsection{A nonlinearity index $\xi$}
To complement the irreversibility analysis of musical compositions, in a second step we consider the temporal arrangement of note sequences within each piece. While it has been extensively certified that music evidences long-range temporal correlations with (typically) a heavy tailed power spectrum \cite{Voss1, pnas_ritmo, Telesca2012, Dagdug2007}, less is known about nonlinear correlations. However, recent studies have reported evidence of nonlinear correlations in musical pieces and discussed on their possible relevance in their structure \cite{gustavo1}.\\
In order to assess the amount of nonlinear temporal correlations, we define a {\it nonlinearity index} $\xi$ inspired in the index previously introduced in \cite{ASHKENAZY200319} by computing the significance (and the amount of nonlinearity) in the Magnitude Detrended Fluctuation Analysis (MDFA) of each musical piece and its Fourier-fixed surrogates (null model with linear correlations). Basically for the calculation of $\xi$ we compare both MDFA computations (original and surrogates) in terms of the local slopes of a fitting polynomial ($\hat{y}$) for the function $F(s)/s$ (see Appendix B for details):
\begin{equation}
\xi = \frac{1}{N_{ws}} \sum_{i=1}^{N_{ws}} \frac{|\hat{y}'(x_i) - \langle \hat{y}'_s(x_i) \rangle_{sur}|}{\sigma(\hat{y}'_s(x_i))_{sur}},
\label{eq:nlinix}
\end{equation}
where $N_{ws}$ is the total number of windows of size $s$, $\hat{y}'(x_i)$ is the first derivative of the polynomial evaluated at the $i$th window of size ($x_i=\log(s_i)$) and  $\langle \hat{y}'_s(x_i) \rangle_{sur}$ and $\sigma(\hat{y}'_s(x_i))_{sur}$ represent the mean and variance of the slopes in the ensemble of surrogates at the $i$th window size respectively (see Appendix B for more details). Surrogates were generated with the Iterative Amplitude Adjusted Fourier Transform (IAAFT) algorithm \cite{SCHREIBER2000346,TSSur}, preserving the marginal distribution and the power spectrum of the original piece.\\
By construction, $\xi \leq 1$ would indicate that the signal only evidences (at most) linear correlations, whereas if $\xi > 1$ then the signal has correlations of nonlinear nature (not reflected in the power spectrum), and the larger $\xi$ the stronger they are \cite{SCHREIBER2000346}. In Figure \ref{fig:nonlin} we measure the nonlinearity index $\xi$ for all the pieces considered in the database. The left panel displays its frequency histogram, certifying that indeed a large majority of pieces display a high nonlinearity index. In the right panel of the same figure we plot $\xi$ as a function of the piece composer's date of birth. Notably, we find that a substantial amount of all musical compositions considered display different degrees of nonlinearity, and the similarity of this panel with panels (a) and (d) in Figure \ref{fig:date} is suggestive. Note that though evidence for different profiles of nonlinear correlations in music scores has been reported previously \cite{gustavo1}, a specific index such as $\xi$, able  to quantify the amount of nonlinearity in a signal, was lacking.

\subsection{Irreversibility vs nonlinearity}
\begin{figure}[htb]
\centering
\includegraphics[width=0.31\columnwidth]{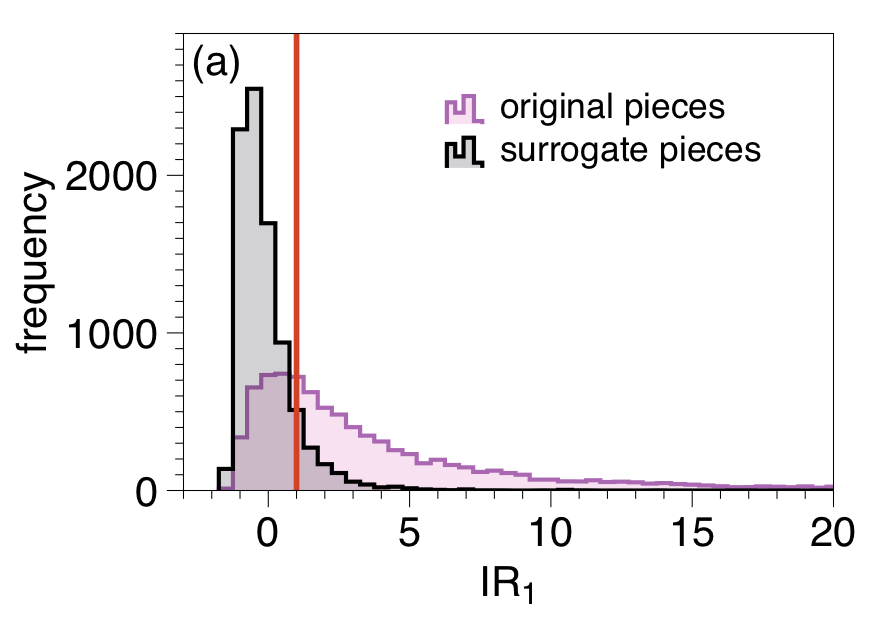}
\includegraphics[width=0.31\columnwidth]{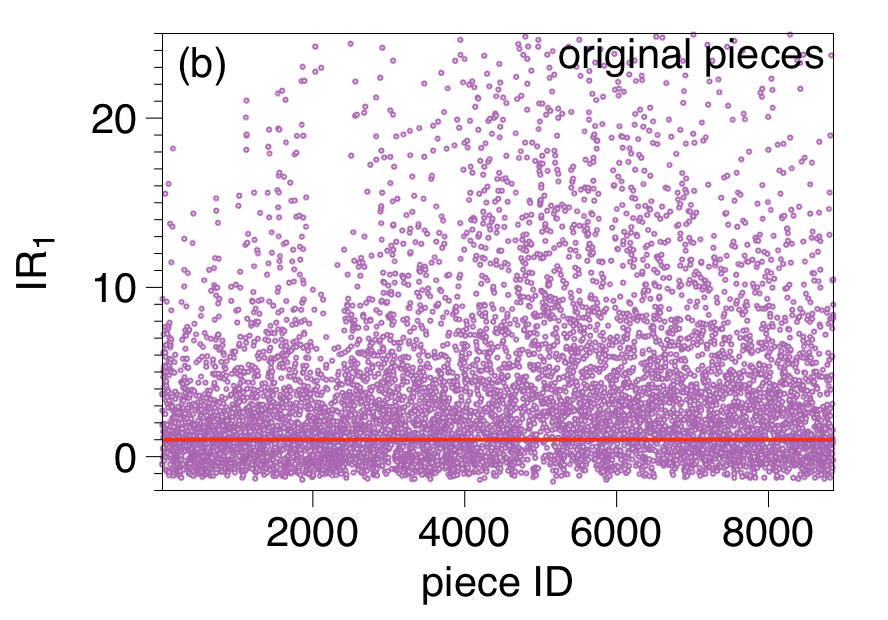}
\includegraphics[width=0.31\columnwidth]{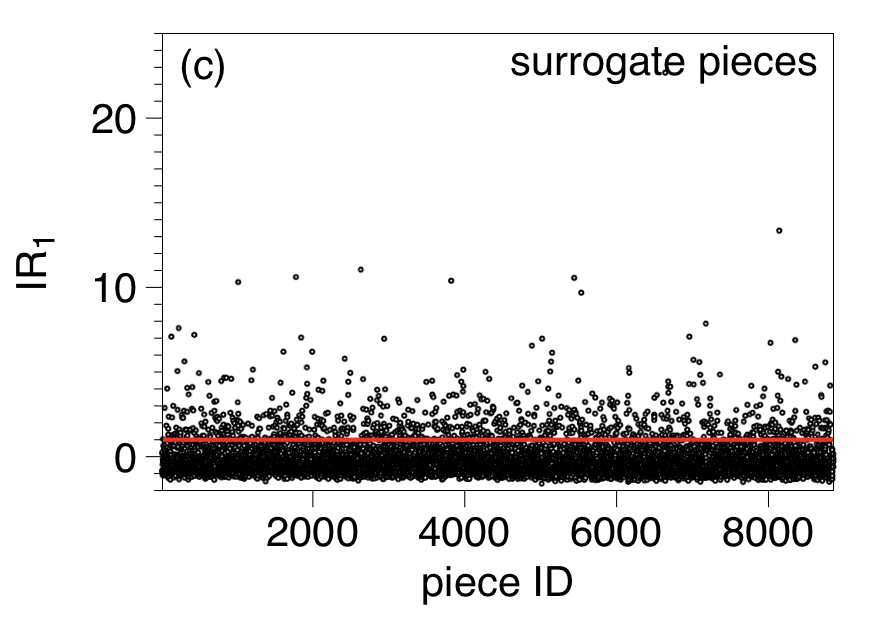}
\caption{{\bf Irreversibility is closely related to nonlinearity. }({\bf a}) Histogram of IR$_1$ for all pieces considered, compared to a similar histogram applied on surrogate pieces where the linear correlation structure (power spectrum) is maintained and other (nonlinear) temporal correlations are removed. We observe that a very large percentage of the surrogate pieces are time reversible, in agreement with the theory. ({\bf b,c}) IR$_1$ of each of the 8856 pieces considered, where we clearly see that many of these are confidently irreversible. Their surrogates, where only linear temporal correlations are maintained, are however largely reversible: the construction of the surrogates removes in most of the cases the irreversible character of the original pieces, further demonstrating that statistical irreversibility is a property that cannot be explained by linear correlations (power spectrum) and ought to be related to nonlinear traits.}
\label{fig:surro_m1}
\end{figure}

In order to understand and link the concept of time irreversibility to nonlinearity, we have investigated to which extent the irreversible character holds when the pieces keep their linear correlations structure but are randomized otherwise. For each piece we have therefore constructed a surrogate piece where the linear correlation structure (power spectrum) is maintained by applying the same technique to the one described previously (IAAFT surrogates), and we then compared IR$_1$ in both cases. In the left panel of figure \ref{fig:surro_m1} we compare the histograms of IR$_1$ for all pieces and for all surrogate pieces. Interestingly, in the case of the surrogates a large percentage of the pieces now display HVG-reversibility (see the middle and right panel of the same figure for a graphical demonstration of the irreversibility loss induced by surrogating the signals). This result can be explained as follows: Gaussian {\it linear} processes with a prescribed power spectrum are indeed reversible. We can understand surrogate pieces as stochastic processes with a prescribed power spectrum (the same as the original piece) but no additional temporal correlation kernels beyond the linear one, and therefore by construction surrogates should be time reversible. All in all, these results point to the fact that time series irreversibility has a connection to nonlinearity.\\

\begin{figure}[h]
\centering
\includegraphics[width=0.35\columnwidth]{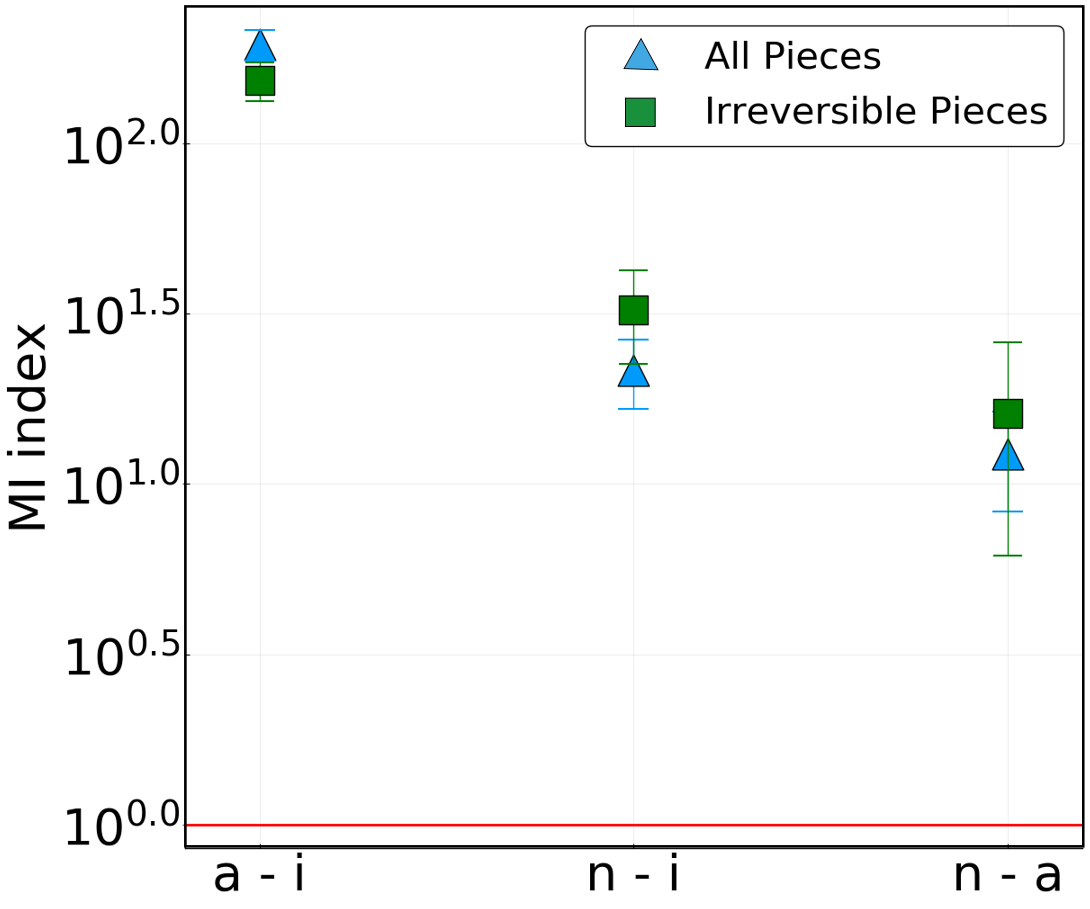}
\caption{{\bf Mutual Information index $\text{MI}_{index}$ between irreversibility, nonlinearity and asymmetry.} Significance for the mutual information between the three properties: Irreversibility ($\text{KLD}_1(\text{in}||\text{out})$) as defined in equation \ref{kld}, nonlinearity ($\langle \hat{y}'(x)\rangle$) defined in the equation \ref{eq:slopes} and asymmetry ($D_{\uparrow \downarrow}$) as defined in the equation \ref{DI}. The red solid line determines the threshold above which there is a statistically significant correlation between indices (the larger $\text{MI}_{index}$, the stronger).
On the x-axis: (a-i) Asymmetry vs Irreversibility, (n-i) Nonlinearity vs Irreversibility and (n-a) Nonlinearity vs Asymmetry. Triangles represent the MI significance for all the pieces, while squares display the same quantities computed only considering the pieces which are confidently irreversible ($\text{IR}_1>1$). The error bars represent the 95\% confidence intervals computed by bootstrapping with replacement.
}
\label{fig:mutuali}
\end{figure}

\noindent Such connection is reinforced by the similarity between the right panel of Figure \ref{fig:nonlin} (nonlinearity index $\xi$) and panels (a) and (d) of Figure \ref{fig:date} where we display the values of IR$_1$ and $\text{KLD}_1(\text{in}||\text{out})$ for all pieces as a function of the date of birth of the composer of the piece. Since the method we use to estimate the amount and significance of nonlinearity has been proved to depend on the length of the series \cite{THEILER199277} (see appendix B), for the estimation of the statistical dependence between nonlinearity and any other property we use the mean value for the local slopes in the MDFA function, which is also related with the amount of nonlinearity and is less dependent of the length of the series \cite{ASHKENAZY200319} (see Appendix B):
\begin{equation}
\langle \hat{y}'(x) \rangle = \frac{1}{N_{ws}} \sum_{i=1}^{N_{ws}} \hat{y}'(x_i),
\label{eq:slopes}
\end{equation}
where $\hat{y}(x)$ is the polynomial fitted for $\log(F(s)/s)$ used for the calculation of the index $\xi$ (quantities are determined before the significance test in equation \ref{eq:nlinix}). In order to quantify the apparent correlation between nonlinearity and irreversibility, we have computed the mutual information between the amount of nonlinearity proxy $\langle \hat{y}'(x)\rangle$ and $\text{KLD}_1(\text{in}||\text{out})$ for all pieces and also for the subset of irreversible pieces, using the discrete mutual information between two random variables $X,Y$
\begin{equation}
\text{MI}(X;Y) = \sum_{y \in {\cal Y}} \sum_{x \in {\cal X}} p(x,y) \log\left(\frac{p(x,y)}{p(x)p(y)} \right),
\end{equation}
we define a mutual information confidence index, in the same fashion as nonlinearity and irreversibility indexes, by substracting the average $\langle \text{MI}\rangle_{null}$ and dividing by the standard deviation $\sigma(\text{MI})_{null}$ of a null model:  \\
\begin{equation}
 \text{MI}_{index} = \frac{\text{MI} - \langle \text{MI} \rangle _{null}} {\sigma(\text{MI})_{null}},
\end{equation}
where the null model is generated random shuffling the elements of one of the variables. The MI indexes are computed by sampling 1000 realizations of the null model. We chose this statistical dependence measure because it captures all kind of correlations (linear and nonlinear) that would be relevant for this case, in contrast with other measures that assume linear dependencies (e.g. Pearson or Spearman coefficients).\\
Results of  $\text{MI}_{index}$ are plotted in figure \ref{fig:mutuali}. We conclude that HVG-irreversibility is indeed intimately related in musical compositions with the presence of nonlinear correlations in the signal. Noting that HVG-reversibility is a proxy for entropy production, this relation manifests a link between a physical concept (dissipation and entropy production) and a statistical one (nonlinear temporal correlations), hence giving a physical interpretation to the latter.

\subsection{Interval asymmetry}
With the aim of linking the patterns observed in terms of irreversibility and nonlinearity with a quantity of musical significance, we finally consider the statistics of {\it intervals}. An interval is defined as the distance between two consecutive notes in a musical piece. For a given sequence of $N$ notes $(n_t)_{t=1}^N$, one can define its respective sequence of intervals as $(i_t)_{t=1}^{N-1}$, with $i_t = n_{t+1} - n_t$, whose properties have been previously analysed \cite{Cocho2008,Cocho2009,Zanette2006,Useche2019,Niklasson2015}.
One particular known result is that small intervals are predominantly {\it descending} while large ones are typically {\it ascending} \cite{VosTroost}. This property introduces an asymmetry in the distribution of intervals which, intuitively, would contribute to the heterogeneity of the joint distributions of consecutive notes in melodic sequences. In order to investigate the relation between interval asymmetry, irreversibility and nonlinearity, we first look at the interval distribution of the set of pieces we studied. We only consider intervals shorter or equal to an octave (12 semitones) since larger intervals are less frequent.
\begin{figure}[htb]
    \centering
    \includegraphics[width=0.35\textwidth]{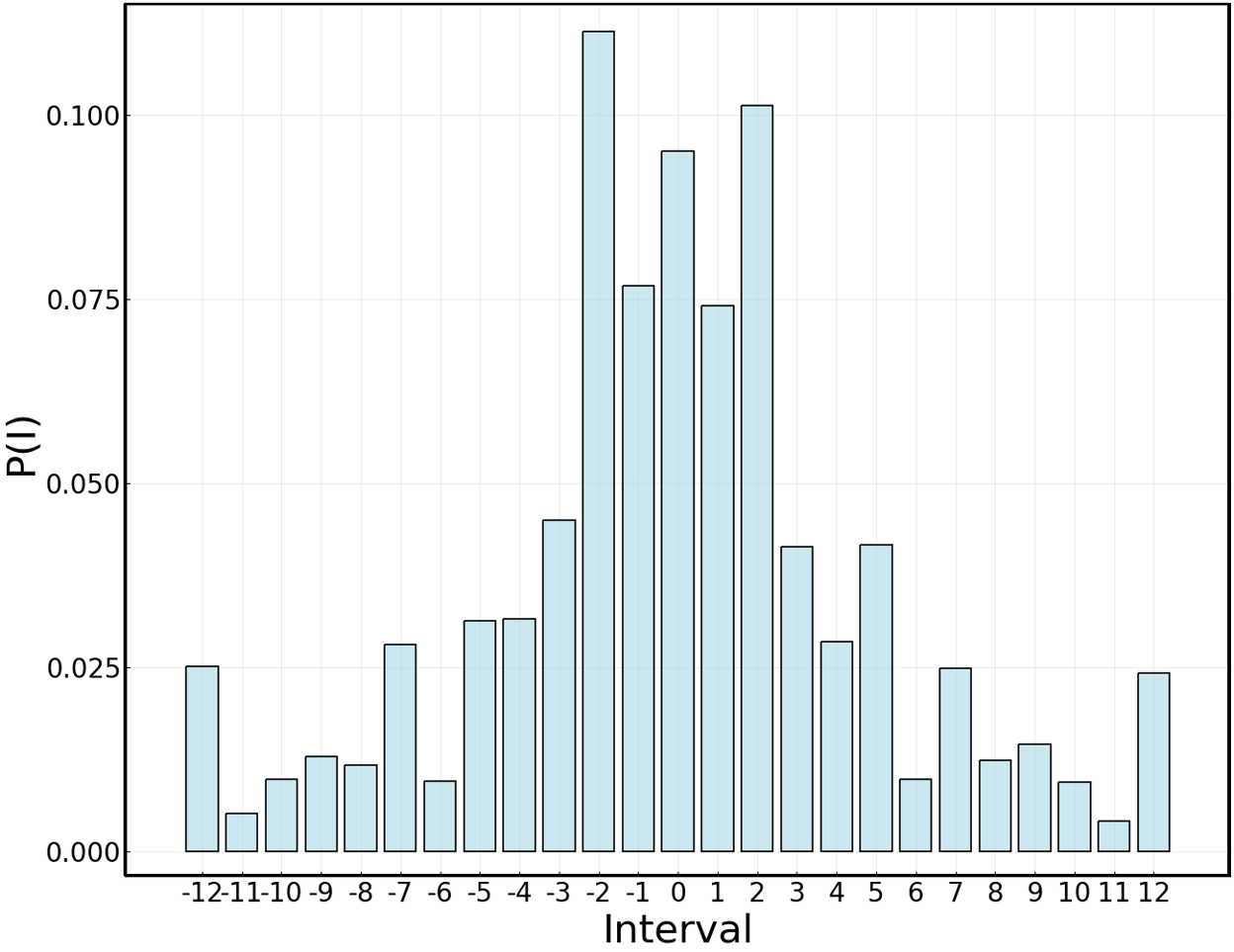}
    \includegraphics[width=0.35\columnwidth]{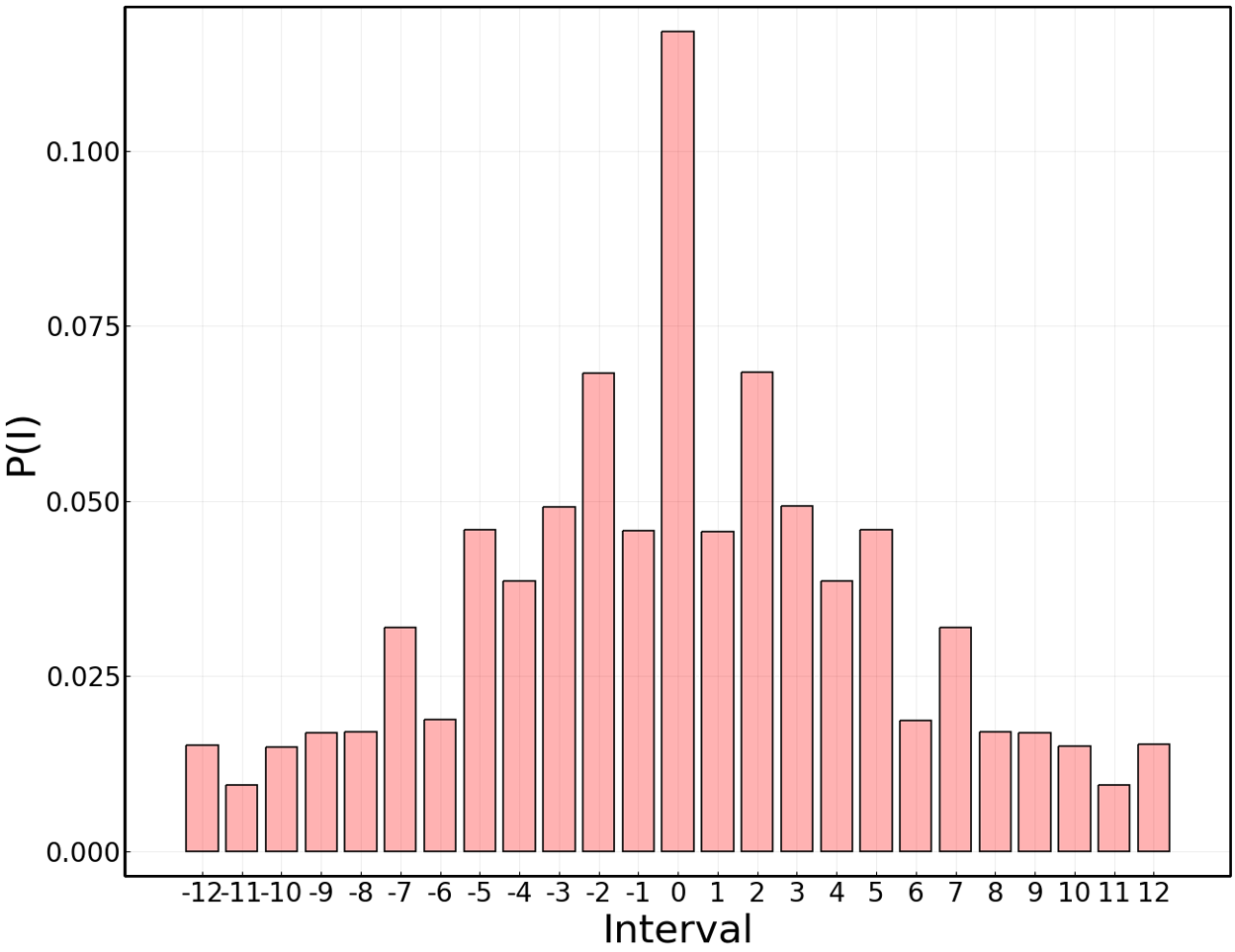}
    \caption{{\bf Interval distributions.} (Left panel) Distribution of the intervals for the all the pieces in this study.  (Right panel) Distribution of the intervals for the ensemble of surrogates constructed in the nonlinearity test.}
    \label{fig:intdists}
\end{figure}
Figure \ref{fig:intdists} displays the interval distributions observed in the complete set of original pieces (left panel) and for the ensemble of surrogates generated previously in the nonlinearity test (right panel). The most evident difference between both distributions is the frequency of the zero interval (when the note keeps the same value), which is lower for the original pieces.
The claim of the difference between ascending and descending intervals holds in our data, the inverals $-4,-3,-2,-1$ are more frequent than $4,3,2,1$ respectively, whereas the interval $5$ is more frequent than $-5$. However, it is not clear that for larger intervals the claim holds. Since the interval distributions are computed for the whole corpus (over 8000 pieces) and not for individual pieces we cannot relate directly this distribution asymmetry with irreversibility. To explore the interval statistics on individual pieces and the possible relation of the asymmetry with nonlinearity and irreversibility we measure the difference between positive (ascending) and negative (descending) intervals ($D_{\uparrow \downarrow}$) for a given piece:
\begin{equation}
    D_{\uparrow \downarrow} =  \frac{ |I_\uparrow  - I_\downarrow |}{ I_\uparrow + I_\downarrow},
    \label{DI}
\end{equation}
where $I_\downarrow$ is the number of negative intervals (when $i_t < 0$) and $I_\uparrow$ the number of positive ones ($i_t > 0$). If the interval distribution of a piece is symmetric (same number of ascending and descending intervals) then the difference $D_{\uparrow \downarrow} = 0$ and if there is only one direction in the melody (ascending or descending) the difference would be $D_{\uparrow \downarrow} = 1$.
\begin{figure}[h]
    \centering
    \includegraphics[width=0.38\columnwidth]{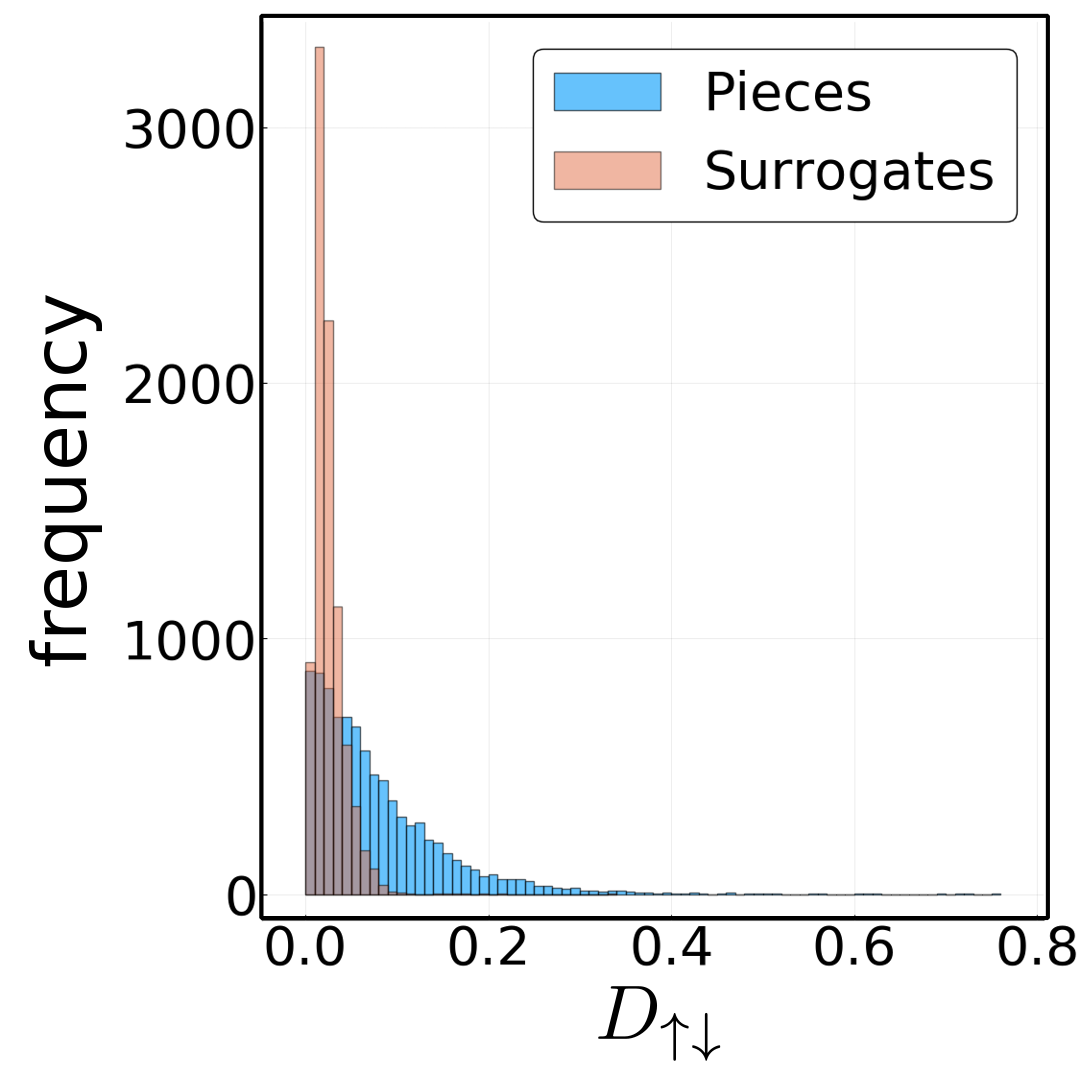}
    \includegraphics[width=0.40\columnwidth]{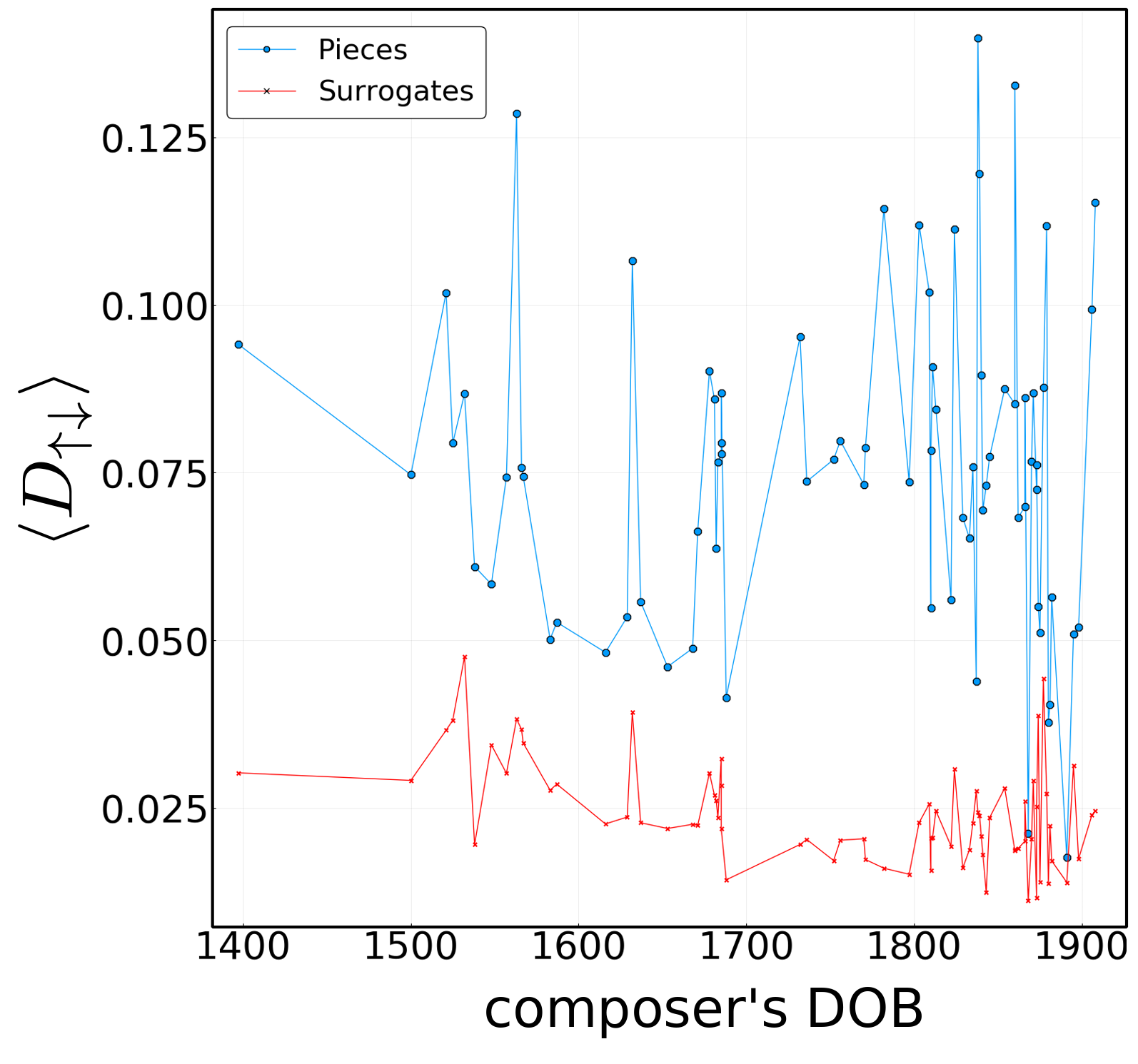}
    \caption{{\bf Interval difference.} (Left panel) Distributions of the values of $D_{\uparrow \downarrow}$ for all the pieces in the study. (Right panel) Mean value of the interval difference for each composer.}
    \label{fig:diffint}
\end{figure}

Results for the interval difference ($D_{\uparrow \downarrow}$) are shown in fig \ref{fig:diffint}, where $D_{\uparrow \downarrow}$ is determined for all the original pieces and for their ensemble of surrogates.
First panel (left) shows the distributions for the values of interval difference ($D_{\uparrow \downarrow}$), the second panel (right) is a plot of the average value of $D_{\uparrow \downarrow}$ for each composer. We systematically find a strong interval assymmetry in a large part of the musical pieces. According to the Mutual Information index (figure \ref{fig:mutuali}) $D_{\uparrow \downarrow}$ indeed strongly correlates with the HVG-irreversibility metric KLD$_1(\text{in}||\text{out})$, and to a lesser extent with the nonlinearity metric $\langle \hat{y}'(x)\rangle$, thus concluding that the irreversibility and nonlinearity traits observed in musical compositions can indeed be narrowed down to musical concepts.

 \section{Discussion}
In this work we have made use of tools from statistical physics, nonlinear dynamics and graph theory to characterise music scores beyond the linear correlation paradigm provided by the power spectrum, in terms of time irreversibility, nonlinear temporal correlations and asymmetric distribution of note intervals. All of these properties intervene in what could be called a ``musical narrative", the flow of a composition. We have established different levels of correlations amongst these quantifiers, some of which invite to reflection. For instance, the finding that in musical compositions nonlinear traits show a considerable correlation with irreversibility, which is a natural hallmark of time directionality, unravels the unforeseen notion that nonlinearity is related to a preferential time arrow. By means of our interval asymmetry exploration we have further found evidence which suggests that --at least some of-- the amount of irreversibility present in musical compositions can be explained in the light of such interval asymmetry. Furthermore, its ensuing connection to nonlinearity gives new insight. It is enticing to associate the idea of irreversibility to structure in the scores at multiple time scales, intercalated with bursts of elements of surprise, it may well be that the degree of irreversibility is linked to the balance between these two factors. \\

\noindent Interestingly, the concepts of irreversibility, directionality and their musical implications have been {\it qualitatively} evoked in music theory under multiple forms. For instance, temporal directionality has been established in terms of irreversible relations of before and after \cite{Hastey}. According to Hastey \cite{Hastey}, for some theorists ``directionality arises in the compositions only through our ability to predict the future course of events. In tonal music a leading tone or passing dissonance implies an expected resolution. Such expectations whether realized or not, constitute in our imagination goals toward which the music is directed". On the other hand, irreversibility was one of the main concerns of musicians such as Anton Webern, one of the exponents of atonality and serialism. Similarly, alegoric relations to thermodynamics and information theory have been previously explored by musicians e.g. Iannis Xenakis, father of stochastic music \cite{Xenakis}. Our approach provides a novel quantitative description of these concepts.\\
In other works, the complexity of music has often been related to linear temporal correlations, using $1/f$ noise as a paradigm offering a balance between predictability and surprise and justifying its enjoyment. However, in our study we show that preserving the exact same power spectrum is not enough to preserve properties such as irreversibility (Fig.\ref{fig:surro_m1}), interval asymmetry (Fig.\ref{fig:diffint}) or nonlinearity (by construction). The fact that these properties are pervasive across five centuries of classical music, together with the observation that such emergence is nontrivial (for instance Gaussian linear stochastic processes are indeed statistically time reversible), challenge the traditional link between pleasantness and linear correlations in music. Furthermore, these three properties show to be statistically related, a result that points to a deep relation between directionality, dissipation, and nonlinearity and their possible relations with the pleasantness in music, a dimension that might be instrumental for the study of perception, music appreciation and cognition.\\

\noindent Summing up, while the statistical properties underlying musical compositions have mainly focused on the presence of linear correlations in the signal (again, the $1/f$ noise paradigm), here we show that classical music compositions over five centuries, encompassing pieces from 77 composers from the Renaissance up to the early modern period, display strong nonlinear correlations. We have certified that such nonlinearities are indeed strongly related to an adequate definition of statistical time irreversibility (HVG-irreversibility), able to quantify the statistical arrow of time in stationary and nonstationary signals alike and well-defined to handle short sequences.
Since HVG-reversibility --as quantified by the Kullback-Leibler divergence between the {\it in} and {\it out} order-$m$ degree distributions of the signal's horizontal visibility graphs-- is a proxy for the amount of thermodynamic entropy produced by a physical signal, exploring time irreversibility in musical compositions allows us to quantify the  process of composition in out-of-thermodynamic-equilibrium' terms. We indeed find that over two thirds of the compositions display this signature. Our study of the value of $\text{KLD}_1(\text{in}||\text{out})$, which is independent of the piece size, shows only small variations over musical periods and thus leads us to conclude that this is a common trait of tonal music. \\

\noindent This work should be taken as a first step of a more in depth, inclusive research program. The three main elements of music are rhythm, melody and harmony \cite{dboyden}. While here we have only addressed melody, we expect the integration of rhythm and harmony in our line of research to be enlightening.
The update of our study in order to include the advent of atonal-dodecaphonic, serial, stochastic, concrete and spectral music, amongst others, is another fascinating open challenge.\\ 

\noindent {\bf Acknowledgements -- } A.G-E. thanks Joshua Plotkin for fruitful discussions and relevant comments in the development of this study. G.M-M. thanks Ismael Darszon for sharing his musical insight, Raul Salgado and Hern\'an Larralde for valuable comments. G.M-M. thanks the hospitality of the Ecole Normale Sup\'erieure (Paris) during a sabbatical leave and the financial support from (DGAPA/UNAM) during that period. L.L. acknowledges funding from EPSRC Early Career Fellowship EP/P01660X/1.\\

\noindent {\bf Author contributions -- } AGE, GMM and LL designed the study. AGE processed and curated the data.
AGE and LL developed the research and implemented all methods. All authors discussed, interpreted results, and wrote the paper.

\section*{Appendix A: Correlations among different Irreversibility Ratio orders }
\begin{figure*}[htb]
\centering
\includegraphics[width=0.24\columnwidth]{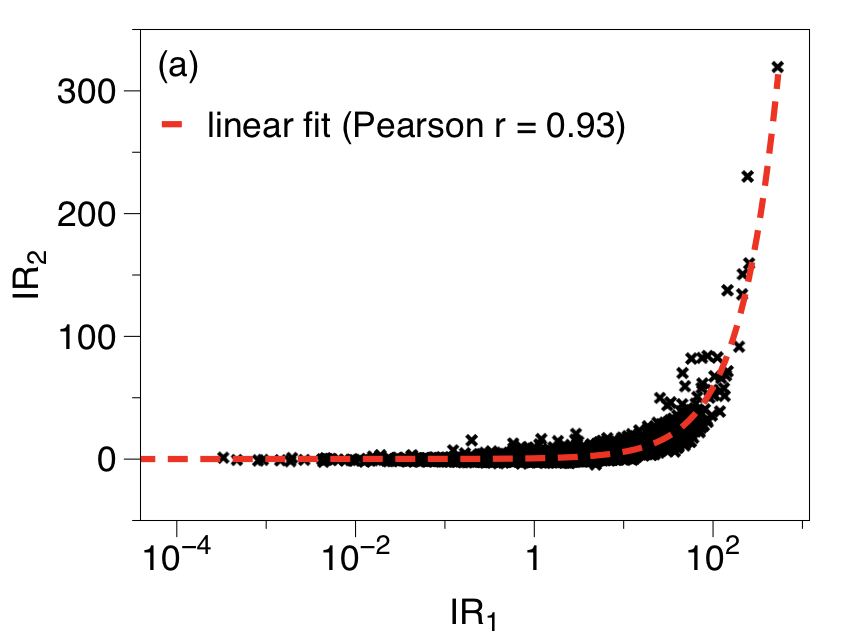}%
\includegraphics[width=0.24\columnwidth]{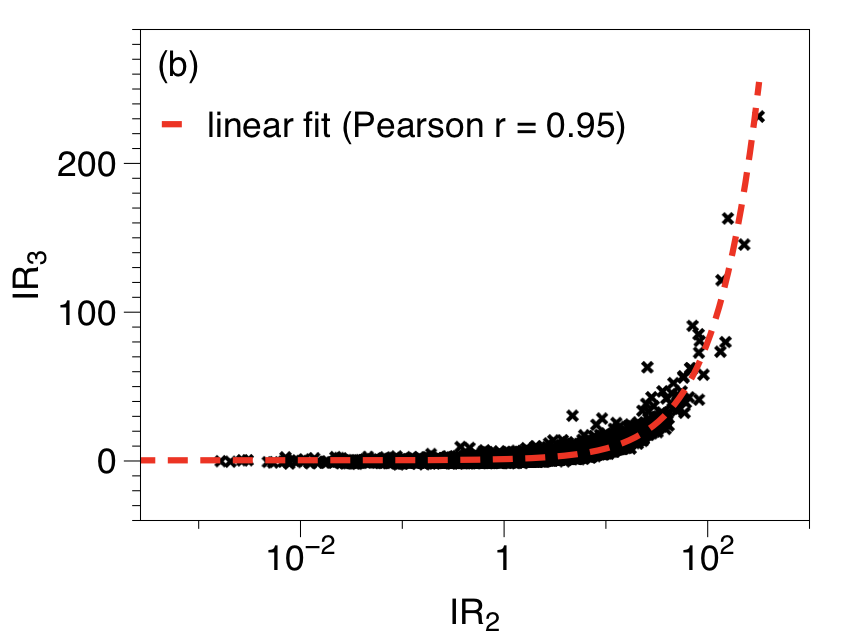}%
\includegraphics[width=0.24\columnwidth]{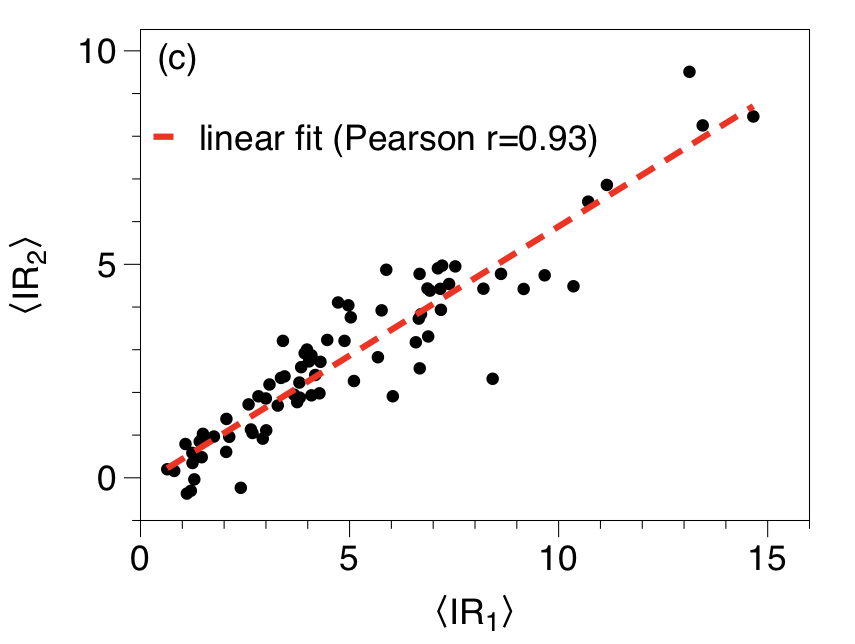}%
\includegraphics[width=0.24\columnwidth]{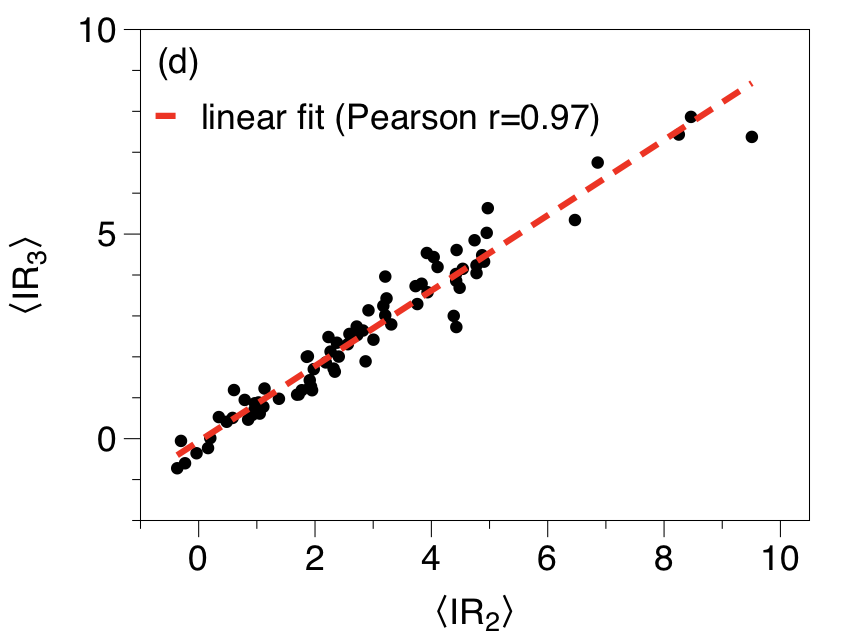}%
\caption{{\bf Irreversibility ratios of different order are correlated. }(a,b) Linear-log scatter plots of Irreversibility ratios IR$_m$ vs IR$_{m-1}$ for each piece in the dataset. The red dashed line is the best linear fit, concluding that all three orders are strongly correlated. (c,d) Linear scatter plots of composer-averaged irreversibility ratios. The strong correlation observed suggests that for the set of signals analysed in this work, IR$_1$ is sufficient to characterise irreversibility and no higher-order quantifiers are needed.}
\label{fig:scatter}
\end{figure*}

In (a) and (b) panels of figure \ref{fig:scatter} we scatter-plot (in linear-log scales, for the sake of illustration) the Irreversibility ratios IR$_m$ vs IR$_{m-1}$ for every piece in the dataset, and compute the Pearson correlation coefficient, finding $r=0.93$ for IR$_2$ vs IR$_1$ and $r=0.95$ for IR$_3$ vs IR$_2$, i.e. very strong correlation. In the (c) and (d) panels of the same figure, a similar analysis is performed (scatter-plots are in linear scales), but we average a single irreversibility ratio per composer (averaging over all the pieces for the same composer), and then correlate $\langle \text{IR}_m \rangle$ vs $\langle \text{IR}_{m-1} \rangle$. Again we find very strong correlation ($r=0.92$ and $0.97$ respectively). Altogether, these results suggest that all three quantities are linearly correlated, meaning that (i) there are no high-order HVG-irreversibility structures in the set of musical compositions analysed, and (ii) accordingly, it is enough to set $m=1$ and safely focus on IR$_1$ to fully account for irreversibility in this context.

\section*{Appendix B: DFA and Magnitude DFA}
The Detrended Fluctuation Analysis (DFA) method, developed by Peng et al. to compute the long range correlations in stationary and nonstationary time series\cite{dfa1}. A brief description of the method is as follows: for a given time series $x(i)$, $i = 1,...,N$, the standard DFA method consists in the following steps: 1) the original signal is integrated $y(j)= \sum_{i=1}^j\left[ x(i) - \langle x \rangle \right]$, where $ \langle x \rangle$ denotes its average value, 2) the integrated time series is then divided into non-overlapping windows of size $s$. 3) Each data segment of length $s$-size is then fitted using a polynomial $y_m(j)$ of degree $m$. 4) Next, the root-mean-square fluctuation from the polynomial, $F(s)$, is calculated:
\begin{equation}
F(s) = \sqrt{\frac{1}{N} \sum_{j=1}^N \left[ y(j) - y_m(j) \right]^2}.
\end{equation}
The procedure is repeated by varying $s$ such that the fluctuation function is obtained in terms of the segment length, which represents the time scale where correlations might be present.
When auto-correlations scale like a power law, the rms fluctuation function $F(s)$ behaves as $F(s) \sim s^\alpha$, where $\alpha$ is the Hurst exponent. A value of $\alpha > 0.5$ indicates the presence of persistent correlations, e.g. $\alpha = 1$ is the case for $1/f$ noise. On the other hand, a value of $0 < \alpha < 0.5$ corresponds to anti-correlations and $\alpha = 0.5$ to white noise
The magnitude Detrended Fluctuation Analysis (MDFA) introduced by Ashkenazy et al\cite{ASHKENAZY200319} is a method capable to detect the presence of nonlinear correlations in a time series.  This method can be summarized by the following recipe: 1) for a given time series $x(i)$ the increment series is defined as $\Delta x(i) \equiv x(i+1) - x(i)$, 2) the increment series is decomposed into a magnitude series and sign series: $\Delta x(i) = sgn(\Delta x(i)) \mid \Delta x(i) \mid$, their respective means are subtracted to avoid artificial trends, 3) because of the limitations of the DFA method for estimating $\alpha < 0.5$ (anti-correlated series), the magnitude and sign series are integrated first to make sure they are positively correlated. 4) The DFA method is implemented on the integrated magnitude and sign series. 5) In order to obtain the respective scaling exponents, the function $F(s)/s$ is estimated, the $1/s$ factor is to compensate the integration made before. If the data obey a scaling law, the fluctuation function should behave as $F(s)/s \sim s^{\alpha - 1}$. It has been shown that the magnitude series ($\mid \Delta x(i) \mid$) is the one that carries information regarding nonlinear correlations in the original time series\cite{Ashkenazy2001}.\\
\begin{figure*}
\centering
\includegraphics[width=0.24\columnwidth]{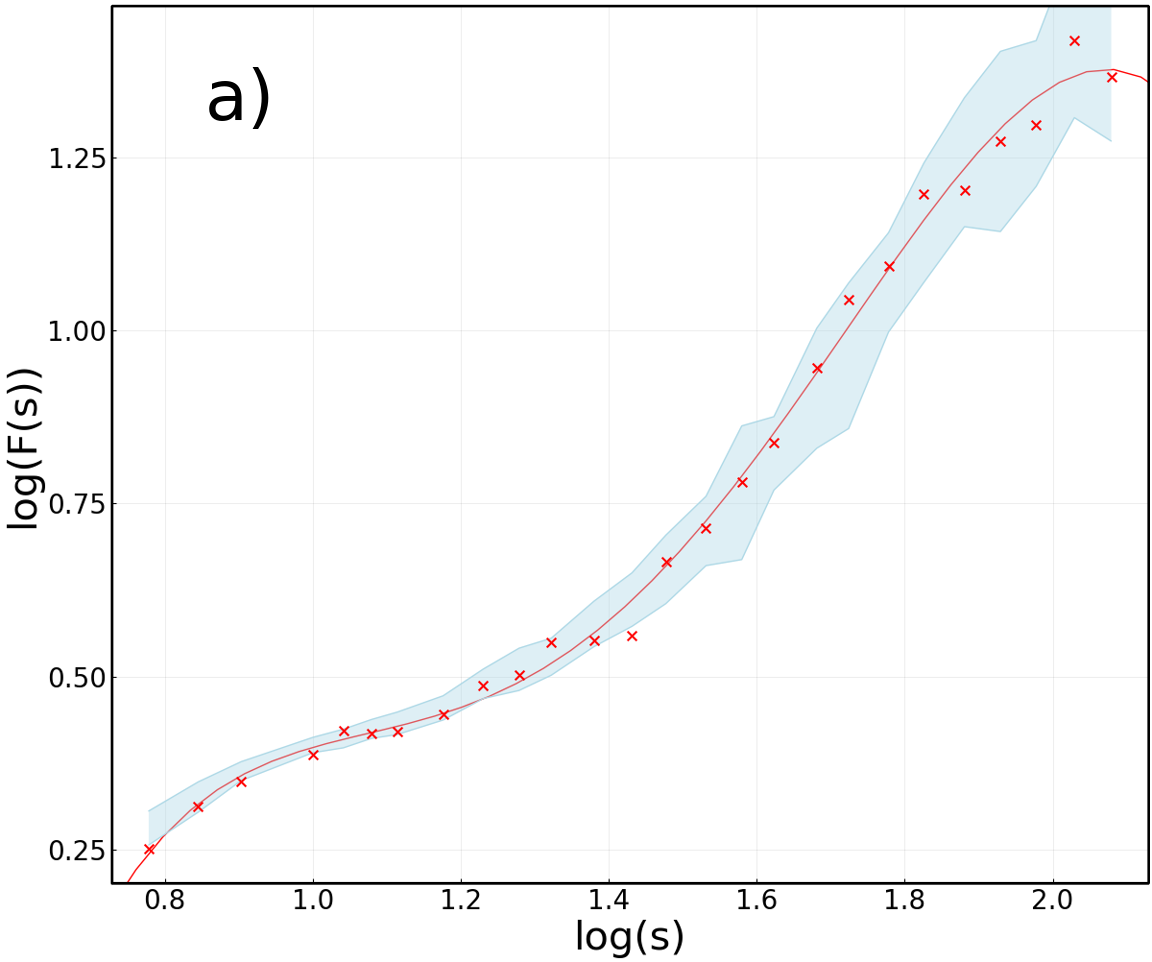}%
\includegraphics[width=0.24\columnwidth]{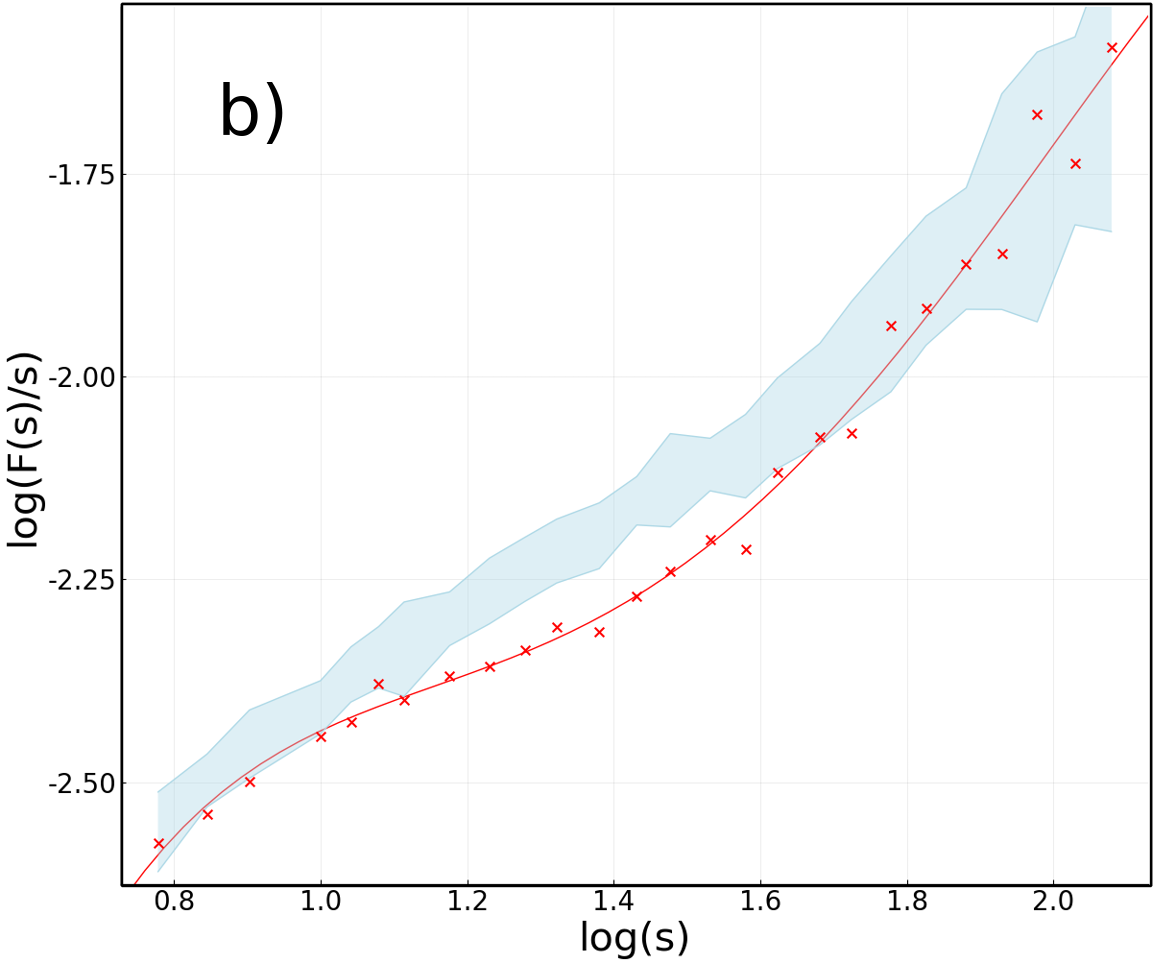}%
\includegraphics[width=0.24\columnwidth]{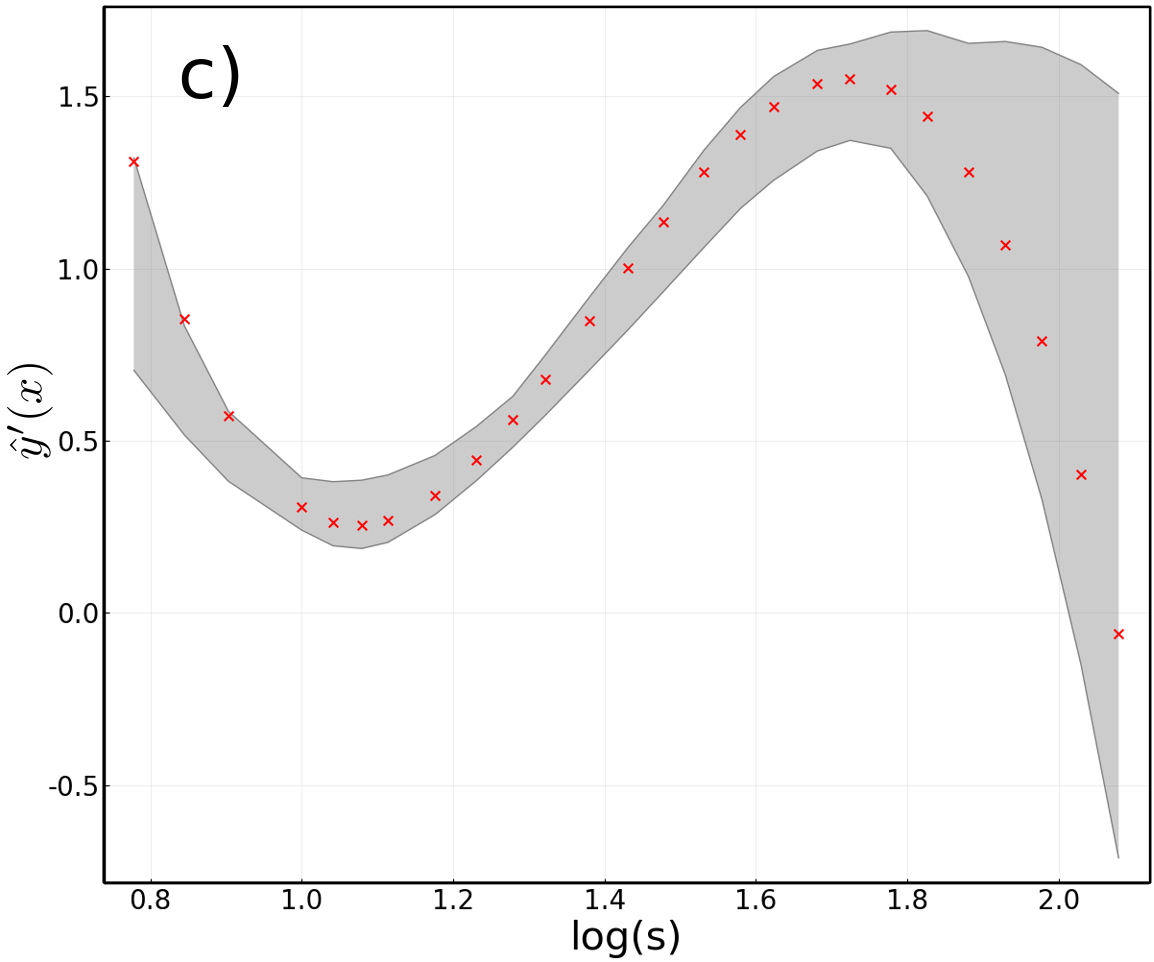}%
\includegraphics[width=0.24\columnwidth]{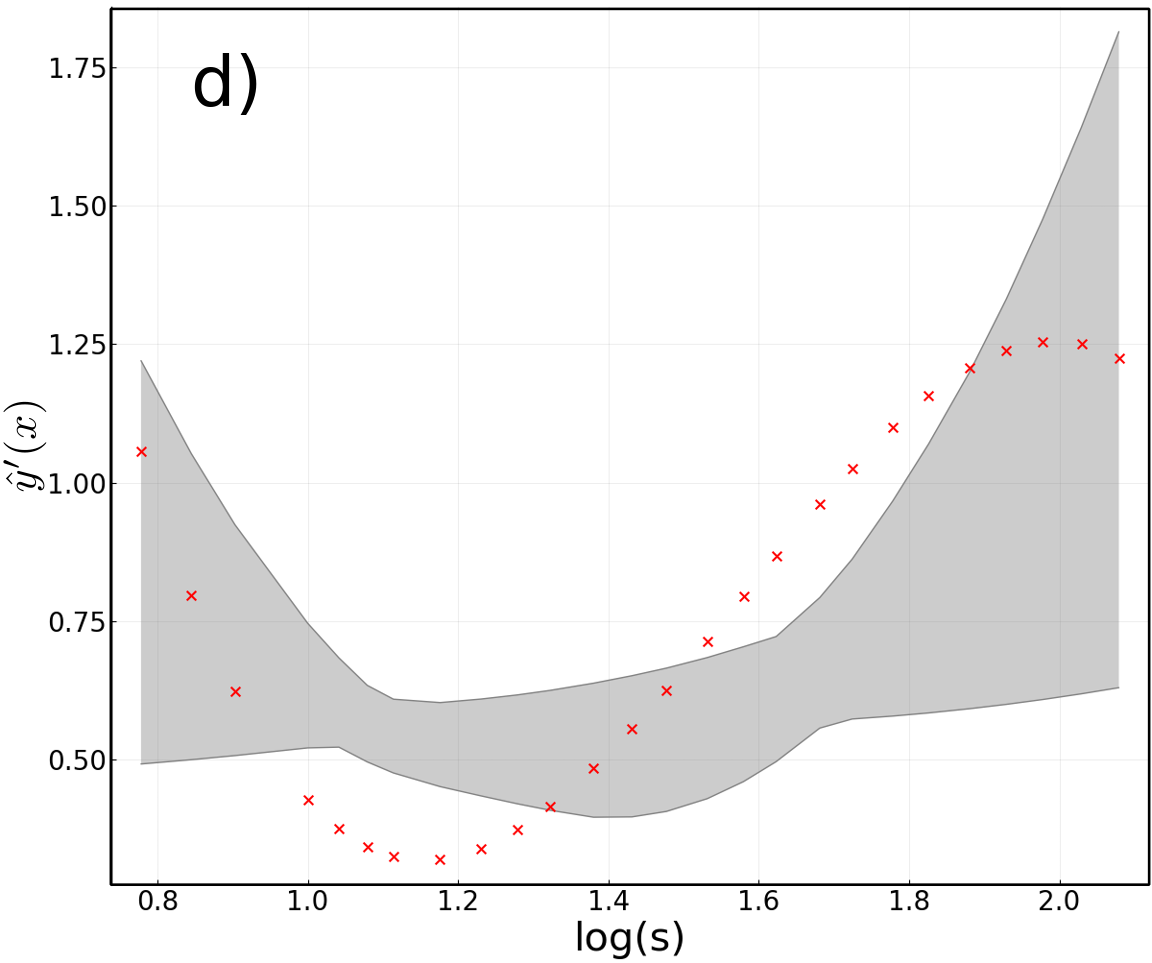}%
\caption{{\bf DFA and MDFA calculations.} One example of a piece from Liszt (excerpt from Christmas tree suite), a) DFA computation with 20 surrogates represented by the shaded area, a polynomial ($\hat{y}$) of order 4 (line) is fitted to the original piece. b)  MDFA computation. c) Slope values for the DFA computation. d) Slope values for the MDFA computation.}
\label{mdfa}
\end{figure*}
To evaluate the evidence and amount of nonlinear correlations it is necessary to compare the MDFA results with the appropriate surrogate data results, these surrogates preserve the linear correlations of the original time series but lack of any possible nonlinear correlations. We generate 20 surrogates for each piece with the Iterative Amplitude Adjusted Fourier Transform (IAAFT) algorithm \cite{SCHREIBER2000346,TSSur}. In panels a and b from figure \ref{mdfa} the results for both original piece (red crosses) and surrogates (shaded area) are shown, the diference of the original data from the surrogates in panel b is evidence for the presence of nonlinear correlations. However, the functions $F(s)$ and $F(s)/s$ do not necessary follow a power law behavior. To be able to quantify the amount of nonlinearity in the time series we define a nonlinearity index $\xi$, given by comparing the scaling behavior of the original data and its surrogates. The proper comparison should be given by the slope of the original data and the slopes of the null model \cite{ASHKENAZY200319}. In our case, instead of having a single scaling we have different regions with different scalings. In order to compare both scaling behaviors we first fit a polynomial $\hat{y}$ to the original MDFA data and to each of its surrogates ($\hat{y}_s$), evaluate the first derivative of $\hat{y}$ (slopes) at each point $(log(s))$ in the MDFA and compute the index $\xi$ as follows:
\begin{equation}
    \xi = \frac{1}{N_{ws}} \sum_{i=1}^{N_{ws}} \frac{|\hat{y}'(x_i) - \langle \hat{y}'_s(x_i) \rangle_{sur}|}{\sigma(\hat{y}'_s(x_i))_{sur}},
\end{equation}
where $N_{ws}$ is the total number of window sizes, $\hat{y}'(x_i)$ is the slope the polynomial evaluated at the $i$th window size ($x_i=log(s_i)$) and  $\langle \hat{y}'_s(x_i) \rangle_{sur}$ and $\sigma(\hat{y}'_s(x_i))_{sur}$ represent the mean and variance of the slopes in the ensemble of surrogates at the $i$th window size respectively. By construction, $\xi \leq 1$ would indicate that the signal only evidences (at most) linear correlations, whereas if $\xi > 1$ the signal has correlations of nonlinear nature (not reflected in the power spectrum), and the larger $\xi$ the stronger.
\begin{figure*}
\centering
\includegraphics[width=0.27\columnwidth]{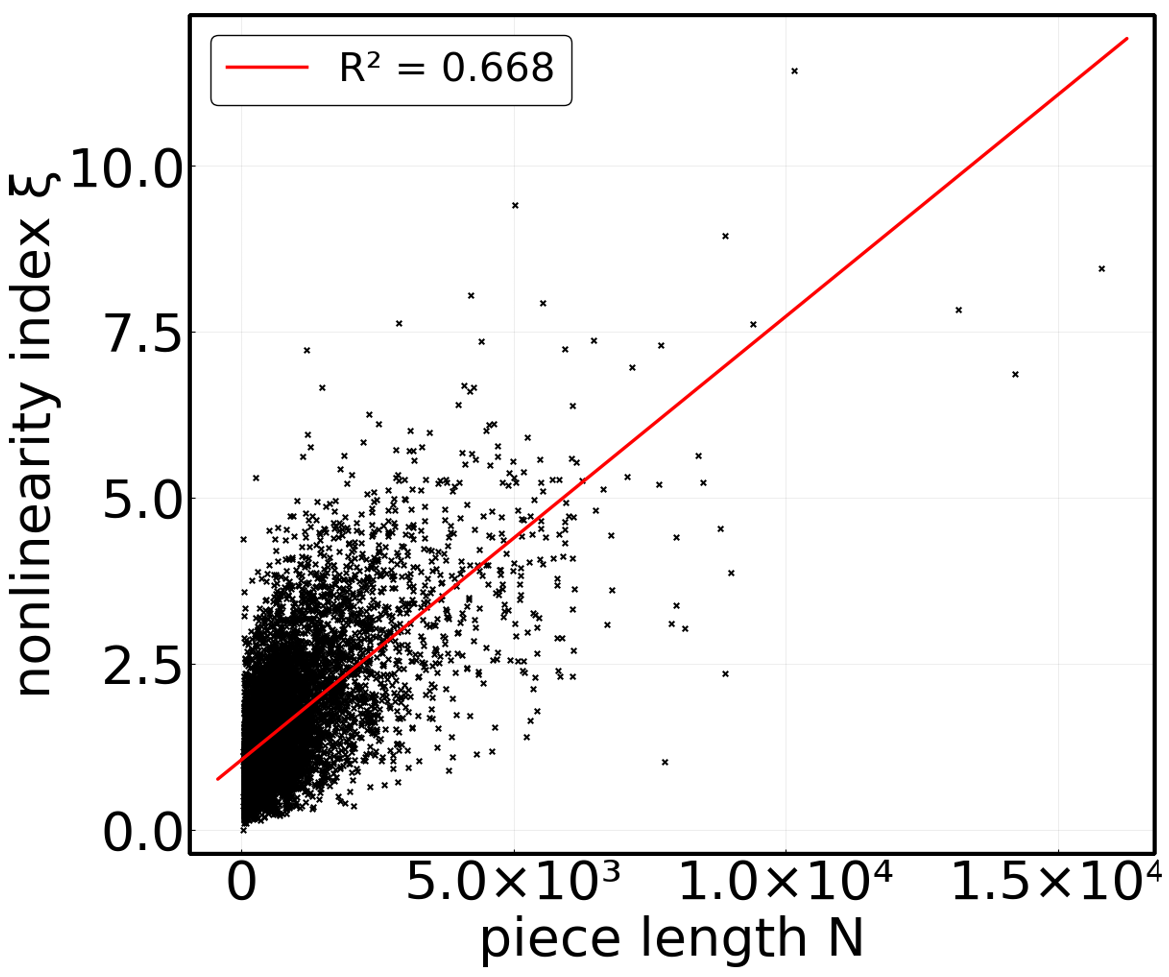}%
\includegraphics[width=0.27\columnwidth]{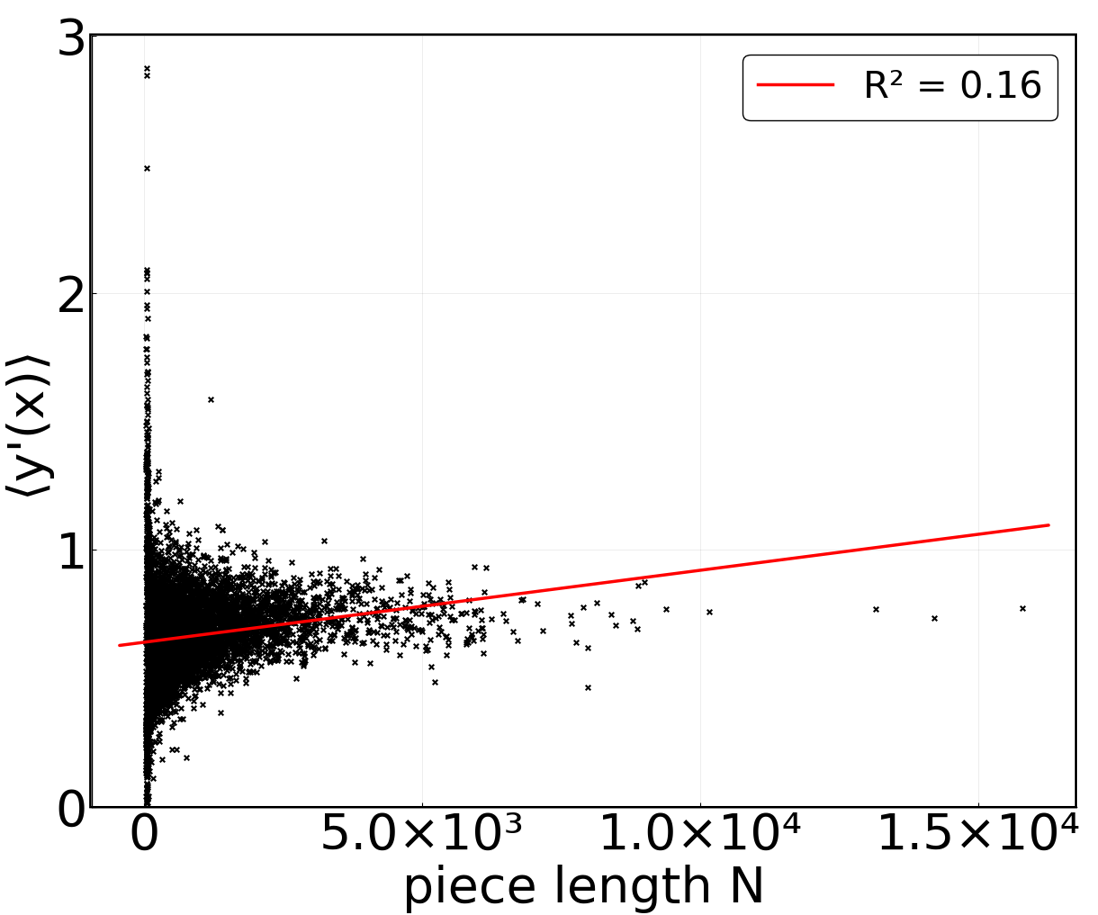}%
\includegraphics[width=0.27\columnwidth]{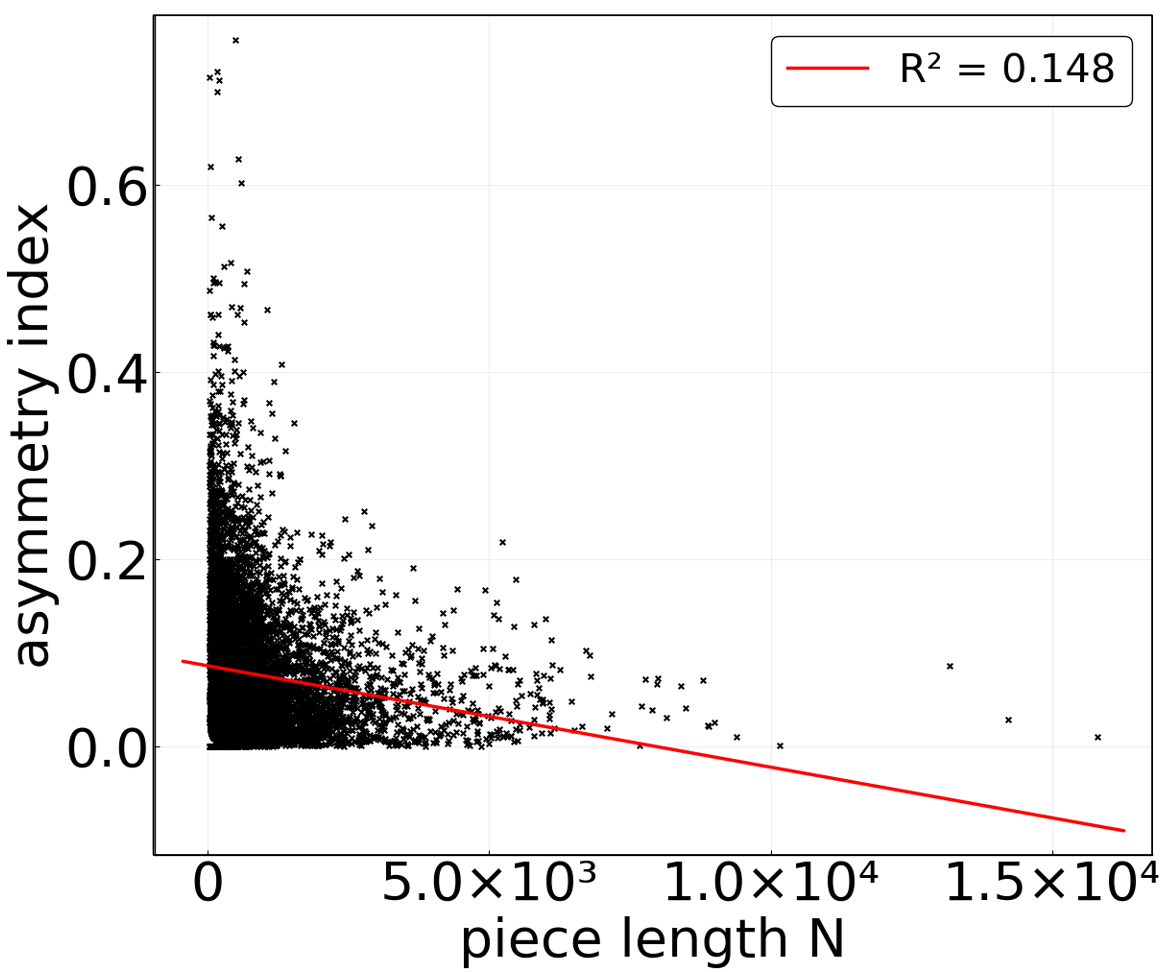}
\caption{{\bf Effect of piece length on the nonlinearity index $\xi$, mean local slope $\langle \hat{y}'(x) \rangle$ and asymmetry $D_{\uparrow \downarrow}$.} }
\label{fig:nasize}
\end{figure*}
\begin{figure*}[htb]
\centering
\includegraphics[width=0.3\columnwidth]{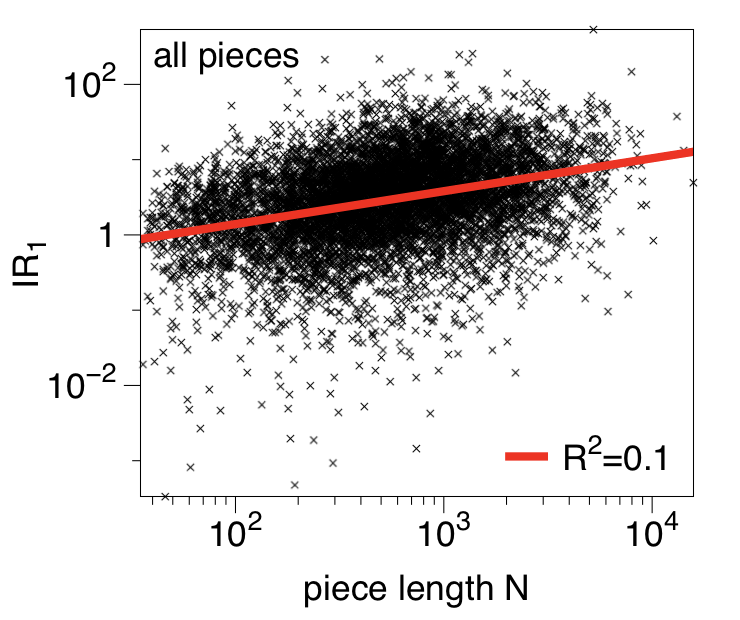}%
\includegraphics[width=0.3\columnwidth]{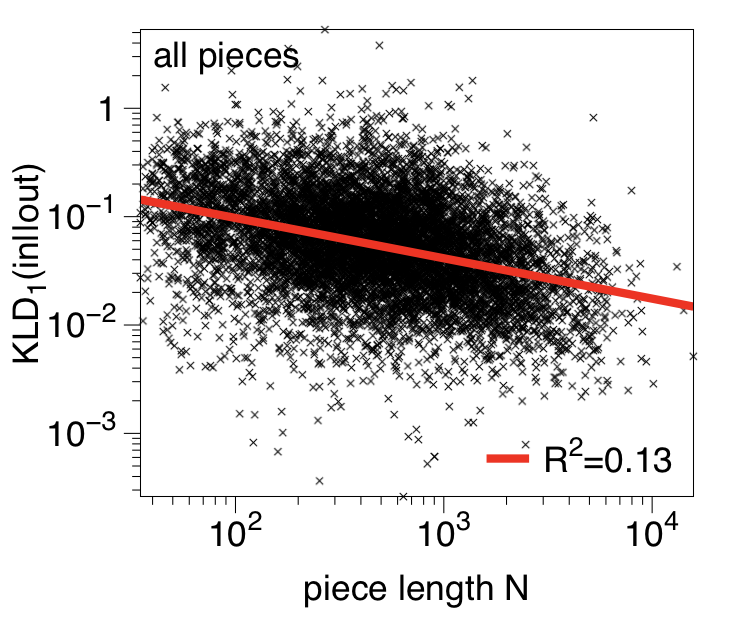}%
\includegraphics[width=0.3\columnwidth]{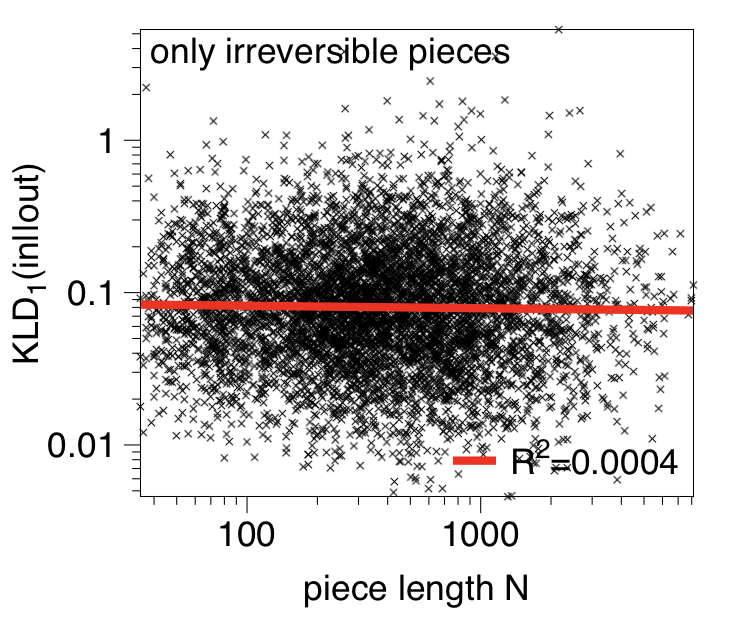}%
\caption{{\bf Effect of piece length on $\text{IR}_1$ and $\text{KLD}_1(\text{in}||\text{out})$} As shown by $R^2$, there are subtle correlations of the irreversibility ratio and irreversibility value with piece length (panels left and middle), and this correlation is removed (right panel) if we only consider the irreversibility value KLD$_1(\text{in}||\text{out})$ the pieces which have been previously certified to be irreversible (IR$_1>1$).}
\label{fig:irsize}
\end{figure*}
\section*{Appendix C: Dependencies of IR$_1$ and KLD$_1(\text{in}||\text{out})$ with series size }
According to the theoretical analysis conducted in Fig.\ref{fig:LOG}, let us assume that a time series of size $N$ is generated by a certain dynamical process. Then we have that
\begin{itemize}
\item if the process is reversible, then IR$_1$ is systematically below 1 and is insensitive to series size $N$ and  on the other hand KLD$_1(\text{in}||\text{out})$ decreases with series size.
\item if the process is irreversible, then KLD$_1(\text{in}||\text{out})$ is reasonably stable and insensitive to series size $N$, whereas IR$_1$ is expected to grow with series size.
\end{itemize}
Let us consider now the musical pieces. Each of them is composed by different composers (so a priori by a possibly different `dynamical process'), and each of them has different length, spanning from dozens to thousands of notes. Comparison across composers and pieces is therefore difficult. In Fig.\ref{fig:irsize} we depict the values of IR$_1$ and KLD$_1(\text{in}||\text{out})$ as a function of the piece length $N$ for all pieces in the dataset (left and middle panels) and only for those pieces that have been certified as irreversible, i.e. those for which IR$_1>1$. Panels are in log-log to have a better visualisation of all the points. We can see in the left panel a  small increasing trend: this is indeed related to all those pieces which are irreversible, whose irreversibility ratio IR$_1$ tends to be larger for larger time series. Similarly, in the middle panel we can appreciate a subtle decreasing trend:
this is indeed related to all those pieces which are reversible, whose reversibility value KLD$_1(\text{in}||\text{out})$ tends to be smaller for larger time series. In the right where panel there is no trend with series length $N$  which is indicative of the size independence of the irreversibility value KLD$_1(\text{in}||\text{out})$.


\end{document}